\documentclass[acmsmall,authorversion]{acmart}

\usepackage{graphicx,calc,array,multirow}
\newlength\myheight
\newlength\mydepth
\settototalheight\myheight{Xygp}
\usepackage{caption}
\usepackage{subcaption}

\AtBeginDocument{%
  \providecommand\BibTeX{{%
    \normalfont B\kern-0.5em{\scshape i\kern-0.25em b}\kern-0.8em\TeX}}}

\acmJournal{PACMHCI}
\acmYear{2023} \acmVolume{7} \acmNumber{CSCW1} \acmArticle{1} \acmMonth{3} \acmPrice{} %

\begin{document}

\title[``Thoughts \& Prayers'' \textit{or} ``\raisebox{-.5\mydepth}{\includegraphics[height=\myheight]{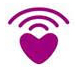}} \&  \raisebox{-.5\mydepth}{\includegraphics[height=\myheight]{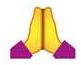}}'']{``Thoughts \& Prayers'' \textit{or} ``\raisebox{-.5\mydepth}{\includegraphics[height=\myheight]{reactions/amp.jpg}} \&  \raisebox{-.5\mydepth}{\includegraphics[height=\myheight]{reactions/pray.jpg}}'': How the Release of New Reactions on CaringBridge Reshapes Supportive Communication During Health Crises}

\author{C. Estelle Smith}
\orcid{0000-0002-4981-7105}
\affiliation{%
  \institution{Colorado School of Mines}
  \city{Golden}
  \state{CO}
  \postcode{80401}
  \country{USA}}
\email{estellesmith@mines.edu}

\author{Hannah Miller Hillberg}
\orcid{0009-0004-4203-7650}
\affiliation{%
  \institution{University of Wisconsin Oshkosh}
  \city{Oshkosh}
  \state{WI}
  \postcode{54901}
  \country{USA}
  }
\email{hillbergh@uwosh.edu}

\author{Zachary Levonian}
\orcid{0000-0002-8932-1489}
\affiliation{%
  \institution{University of Minnesota}
  \city{Minneapolis}
  \state{MN}
  \postcode{55455}
  \country{USA}}
\email{levon003@umn.edu}

\renewcommand{\shortauthors}{C. Estelle Smith, et al.}

\begin{abstract}
Following Facebook's introduction of the ``Like''~\raisebox{-.5\mydepth}{\includegraphics[height=\myheight]{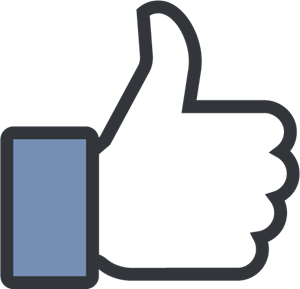}}~in 2009, CaringBridge (a nonprofit health journaling platform) implemented a ``Heart''~\raisebox{-.5\mydepth}{\includegraphics[height=\myheight]{reactions/amp.jpg}} symbol as a single-click reaction affordance in 2012. In 2016, Facebook expanded its Like into a \textit{set} of emotion-based reactions. In 2021, CaringBridge likewise added three new reactions: \raisebox{-.5\mydepth}{\includegraphics[height=\myheight]{reactions/pray.jpg}}~``Prayer'', \raisebox{-.5\mydepth}{\includegraphics[height=\myheight]{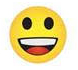}}~``Happy'', and \raisebox{-.5\mydepth}{\includegraphics[height=\myheight]{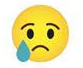}}~``Sad.'' Through user surveys ($N=808$) and interviews ($N=13$), we evaluated this product launch. Unlike Likes on mainstream social media, CaringBridge's single-click Heart was consistently interpreted as a simple, meaningful expression of acknowledgement and support. Although most users \textit{accepted} the new reactions, the product launch transformed user perceptions of the feature and ignited major disagreement regarding the meanings and functions of reactions in the high stakes context of health crises. Some users found the new reactions to be useful, convenient, and reducing of caregiver burden; others felt they cause emotional harms by stripping communication of meaningful expression and authentic care. Overall, these results surface tensions for small social media platforms that need to survive amidst giants, as well as highlighting crucial trade-offs between the cognitive effort, meaningfulness, and efficiency of different forms of Computer-Mediated Communication (CMC). Our work provides three contributions to support researchers and designers in navigating these tensions: (1) empirical knowledge of how users perceived the reactions launch on CaringBridge; (2) design implications for improving health-focused CMC; and (3) concrete questions to guide future research into reactions and health-focused CMC.
\end{abstract}



\newcommand{\participants}{
\begin{table}[]
\label{tab:survey_participants}
\footnotesize
\textbf{Table 1a: SURVEY RESPONDENTS} \\
\phantom{text}

\begin{tabular}{llrrr}
\toprule
 & & All respondents ($n$=808) & Authors ($n$=437) & Visitors ($n$=371) \\ \midrule
\multirow{4}{*}{Role} & Patient & 184 (22.8\%) & 160 (36.6\%) & 24 (6.5\%) \\ 
 & Caregiver & 228 (28.2\%) & 211 (48.3\%) & 17 (4.6\%) \\ 
 & Close Friend/Family & 380 (47.0\%) & 169 (38.7\%) & 211 (56.9\%) \\ 
 & Friend/Acquaintance & 355 (43.9\%) & 125 (28.6\%) & 230 (62.0\%) \\ 
 \midrule
\multirow{3}{*}{Gender} & Female & 691 (85.5\%) & 365 (83.5\%) & 326 (87.9\%) \\ 
 & Male & 97 (12.0\%) & 62 (14.2\%) & 35 (9.4\%) \\ 
 & NB & 20 (2.5\%) & 10 (2.3\%) & 10 (2.7\%) \\ 
 \midrule
\multirow{3}{*}{Age} & 18-24 years old & 16 (2.0\%) & 3 (0.7\%) & 13 (3.5\%) \\ 
 & 25-34 years old & 51 (6.3\%) & 25 (5.7\%) & 26 (7.0\%) \\ 
 & 35-44 years old & 153 (18.9\%) & 94 (21.5\%) & 59 (15.9\%) \\ 
 & 45-54 years old & 186 (23.0\%) & 114 (26.1\%) & 72 (19.4\%) \\ 
 & 55-64 years old & 225 (27.8\%) & 124 (28.4\%) & 101 (27.2\%) \\ 
 & 65-74 years old & 146 (18.1\%) & 65 (14.9\%) & 81 (21.8\%) \\ 
 & 75 years or older & 31 (3.8\%) & 12 (2.7\%) & 19 (5.1\%) \\ 
 \midrule
\multirow{7}{*}{Spirituality} & Christian & 543 (67.2\%) & 299 (68.4\%) & 244 (65.8\%) \\ 
 & Spiritual, Non-Religious & 118 (14.6\%) & 55 (12.6\%) & 63 (17.0\%) \\ 
  & Agnostic & 47 (5.8\%) & 26 (5.9\%) & 21 (5.7\%) \\ 
 & Atheist & 39 (4.8\%) & 18 (4.1\%) & 21 (5.7\%) \\ 
 & Jewish & 35 (4.3\%) & 25 (5.7\%) & 10 (2.7\%) \\ 
 & Muslim & 7 (0.9\%) & 3 (0.7\%) & 4 (1.1\%) \\ 
 & Other/Prefer Not to Say & 37 (4.6\%) & 18 (4.1\%) & 19 (5.1\%) \\ 
 \midrule
\multirow{3}{1.2cm}{More or Fewer Reactions?} & More & 379 (46.9\%) & 207 (47.4\%) & 172 (46.4\%) \\ 
 & Same & 365 (45.2\%) & 199 (45.5\%) & 166 (44.7\%) \\ 
 & Fewer & 64 (7.9\%) & 31 (7.1\%) & 33 (8.9\%) \\ 
 \midrule
\multicolumn{2}{l}{Has seen the new Reactions} & 404 (50.0\%) & 191 (43.7\%) & 213 (57.4\%) \\ 
\multicolumn{2}{l}{Has used the new Reactions} & 237 (29.3\%) & 115 (26.3\%) & 122 (32.9\%) \\ 
\bottomrule
\end{tabular}

\phantom{text} \\
\textbf{Table 1b: INTERVIEW PARTICIPANTS} \\
\phantom{text} \\

\begin{tabular}{llccclllll}
\toprule
 & ID & Patient & Caregiver & Friend & Spirituality & Age & Gender & Seen new? & More or fewer? \\ \toprule

& A1 &  & \checkmark & \checkmark & Christian & 35-44 & F & No & More \\
 & A2 &  &  & \checkmark & Christian & 55-64 & M & Yes & More \\
 Authors & A3 & \checkmark &  & \checkmark & Christian & 55-64 & F & Yes & Same \\
 & A4 &  & \checkmark & \checkmark & Christian & 35-44 & F & Yes & More \\
 & A5 &  & \checkmark & \checkmark & Agnostic & 55-64 & F & No & Same \\
 & A6 & \checkmark & \checkmark & \checkmark & Jewish & 65-74 & F & Yes & More \\
 & A7 &  & \checkmark & \checkmark & Jewish & 35-44 & F & No & Same \\
 \midrule
& V1 &  &  & \checkmark & Christian & 25-34 & F & Yes & Same \\
 & V2 &  &  & \checkmark & Christian & 55-64 & F & No & More \\
 Visitors & V3 &  &  & \checkmark & Christian & 65-74 & F & Yes & More \\
 & V4 &  &  & \checkmark & Atheist & 75+ & F & Yes & Fewer \\  
 & V5 &  &  & \checkmark & Spiritual & 55-64 & M & Yes & Same \\
 & V6 &  &  & \checkmark & Spiritual & 45-54 & F & No & More \\
 \bottomrule
\end{tabular}

\caption{Summary of research participants. Interviewees (Table 1b, bottom) are a strict subset of all survey respondents (Table 1a, top).}
\label{tab:participants}
\end{table}

}

\newcommand{\reactionpicker}{
\begin{figure}%
    \centering
    \includegraphics[width=0.65\textwidth]{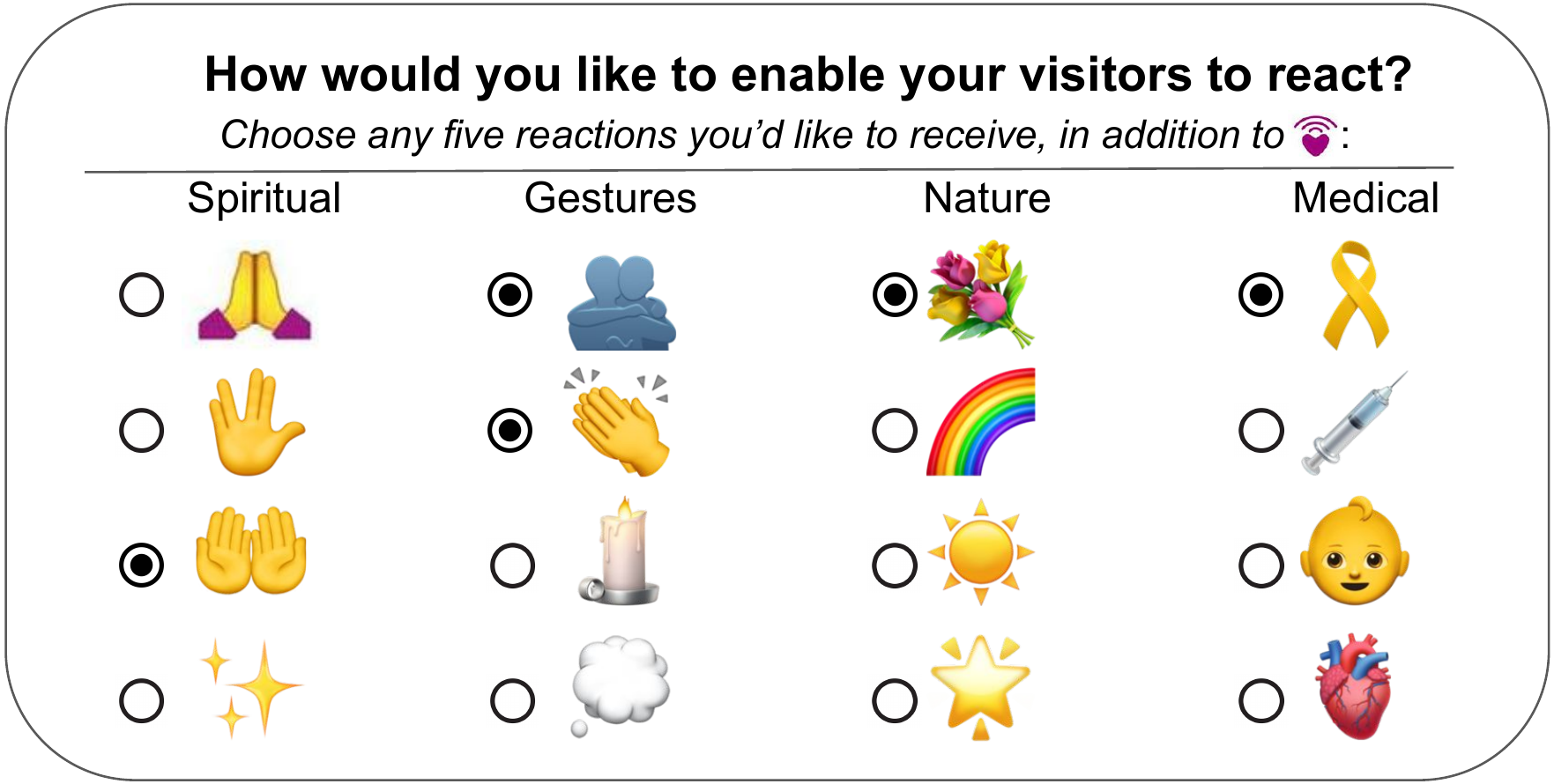}
    \caption{Prospective reaction picker feature. We envision that a selection menu could be designed to include custom-designed and branded reactions graphics, enabling CaringBridge to constrain the set of all possible reactions, while also enabling authors to specify a specific subset of reactions that they most want to receive for their particular health journey.}
    \label{fig:reactionpicker}
\end{figure}
}

\newcommand{\UIappearance}{
\begin{figure}[t]
    \centering
    \begin{subfigure}[t]{0.45\textwidth}
    \centering
        \includegraphics[width=2in]{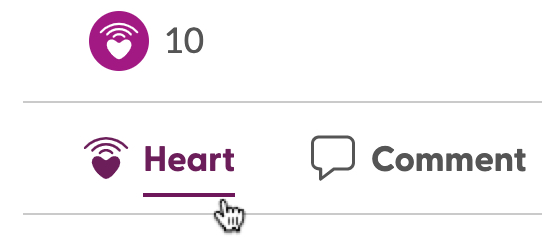}
        \caption{}%
    \end{subfigure}%
    ~ 
    \begin{subfigure}[t]{0.45\textwidth}
    \centering
        \includegraphics[width=1.9in]{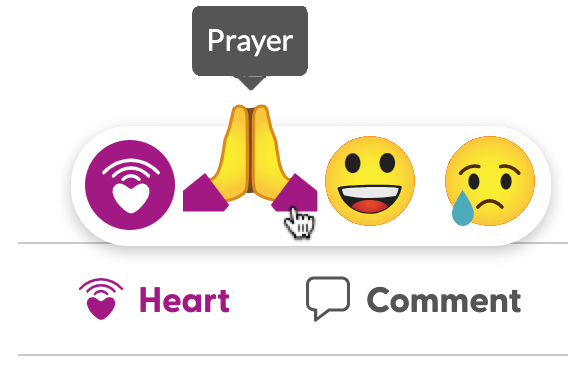}
        \caption{}%
    \end{subfigure}
        \begin{subfigure}[t]{0.45\textwidth}
    \centering
        \includegraphics[width=1.5in]{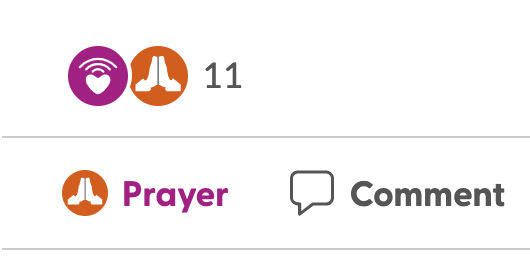}
        \caption{}%
    \end{subfigure}
        \begin{subfigure}[t]{0.45\textwidth}
    \centering
        \includegraphics[width=1.5in]{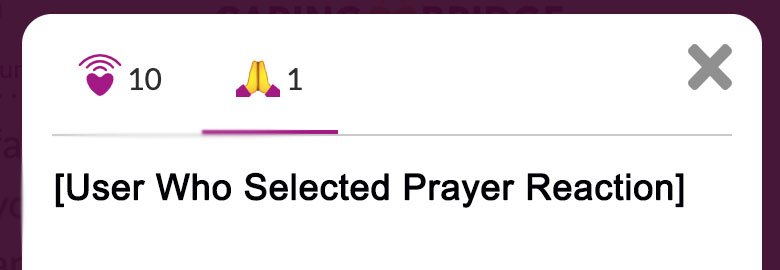}
        \caption{}%
    \end{subfigure}
    \caption{Example UI/UX for New Reactions on CaringBridge. In this example, a user adds a "Prayer" Reaction to a Journal update that had previously received 10 Heart Reactions from others. (a) Users can select the Heart reaction by clicking on "\raisebox{-.5\mydepth}{\includegraphics[height=\myheight]{reactions/amp.jpg}} Heart" at the bottom left corner of a Journal update. Hovering for 250ms causes an additional panel of reaction options to appear. (b) Users can select \raisebox{-.5\mydepth}{\includegraphics[height=\myheight]{reactions/amp.jpg}},  \raisebox{-.5\mydepth}{\includegraphics[height=\myheight]{reactions/pray.jpg}}, \raisebox{-.5\mydepth}{\includegraphics[height=\myheight]{reactions/happy.jpg}} or \raisebox{-.5\mydepth}{\includegraphics[height=\myheight]{reactions/sad.jpg}}. (c) After selecting \raisebox{-.5\mydepth}{\includegraphics[height=\myheight]{reactions/pray.jpg}}, the count of all reactions is incremented, and a Prayer icon appears beside the Heart. (d) Clicking on the count of reactions triggers a pop-up window that shows which users left which reactions.}\label{fig:UI}
\end{figure}
}

\newcommand{\resultssummary}{
\begin{figure}
    \centering
    \includegraphics[width=\textwidth]{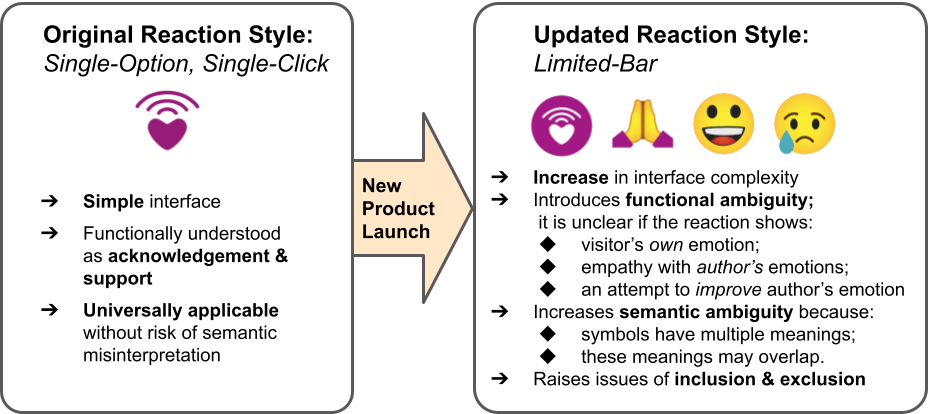}
    \caption{Summary of Major Takeaways from Results. Left: Users' descriptions of the "Single-Option, Single-Click" reaction style. Right: Users' experiences following the new product launch of the "Limited Bar."}
    \label{fig:resultssummary}
\end{figure}
}

\newcommand{\subgroup}{

\begin{table}[]
\label{tab:participants_subgroup}
\begin{tabular}{@{}llrrr@{}}
\toprule
 & & All respondents (\%, $n$=404) & Authors (\%, $n$=191) & Visitors (\%, $n$=213) \\ \midrule
\multirow{4}{*}{Role} & Patient & 84 (20.8\%) & 69 (36.1\%) & 15 (7.0\%) \\ 
 & Caregiver & 92 (22.8\%) & 82 (42.9\%) & 10 (4.7\%) \\ 
 & Close Friend/Family & 192 (47.5\%) & 77 (40.3\%) & 115 (54.0\%) \\ 
 & Casual Friend/Acquaintance & 188 (46.5\%) & 58 (30.4\%) & 130 (61.0\%) \\ 
 \midrule
\multirow{3}{*}{Gender} & Female & 327 (80.9\%) & 150 (78.5\%) & 177 (83.1\%) \\ 
 & Male & 60 (14.9\%) & 34 (17.8\%) & 26 (12.2\%) \\ 
 & NB & 17 (4.2\%) & 7 (3.7\%) & 10 (4.7\%) \\ 
 \midrule
\multirow{3}{*}{Age} & 18-24 years old & 10 (2.5\%) & 0 (0.0\%) & 10 (4.7\%) \\ 
 & 25-34 years old & 25 (6.2\%) & 13 (6.8\%) & 12 (5.6\%) \\ 
 & 35-44 years old & 69 (17.1\%) & 33 (17.3\%) & 36 (16.9\%) \\ 
 & 45-54 years old & 88 (21.8\%) & 48 (25.1\%) & 40 (18.8\%) \\ 
 & 55-64 years old & 115 (28.5\%) & 58 (30.4\%) & 57 (26.8\%) \\ 
 & 65-74 years old & 76 (18.8\%) & 31 (16.2\%) & 45 (21.1\%) \\ 
 & 75 years or older & 21 (5.2\%) & 8 (4.2\%) & 13 (6.1\%) \\ 
 \midrule
\multirow{7}{*}{Spirituality} & Christian & 283 (70.0\%) & 130 (68.1\%) & 153 (71.8\%) \\ 
 & Jewish & 17 (4.2\%) & 14 (7.3\%) & 3 (1.4\%) \\ 
 & Muslim & 5 (1.2\%) & 2 (1.0\%) & 3 (1.4\%) \\ 
 & Spiritual, Non-Religious & 57 (14.1\%) & 26 (13.6\%) & 31 (14.6\%) \\ 
 & Atheist & 12 (3.0\%) & 5 (2.6\%) & 7 (3.3\%) \\ 
 & Agnostic & 22 (5.4\%) & 10 (5.2\%) & 12 (5.6\%) \\ 
 & Prefer Not to Say & 21 (5.2\%) & 7 (3.7\%) & 14 (6.6\%) \\ 
 \midrule
\multirow{3}{1.2cm}{More or Fewer Reactions?} & More & 198 (49.0\%) & 95 (49.7\%) & 103 (48.4\%) \\ 
 & Same & 180 (44.6\%) & 86 (45.0\%) & 94 (44.1\%) \\ 
 & Fewer & 26 (6.4\%) & 10 (5.2\%) & 16 (7.5\%) \\ 
 \midrule
\multicolumn{2}{l}{Has seen the new Reactions} & 404 (100.0\%) & 191 (100.0\%) & 213 (100.0\%) \\ 
\multicolumn{2}{l}{Has used the new Reactions} & 237 (58.7\%) & 115 (60.2\%) & 122 (57.3\%) \\ 
\bottomrule
\end{tabular}
\caption{Table 1, but just the people who have seen.}
\end{table}

}

\maketitle

\section{Introduction}
Online health communities (OHCs)--defined as online social networks (OSNs) related to health--are pivotal resources for millions of people facing critical or life-threatening health conditions. OHCs can be situated either upon niche, health-focused platforms, or nestled within larger mainstream OSNs. In both cases, patients and caregivers engage in Computer-Mediated Communication (CMC) by sharing sensitive disclosures in support-seeking posts, and getting spiritual~\cite{smith_i_2020,smith_what_2021} and social support~\cite{cutrona_controllability_1992,uchino_social_2004,macleod_be_2017,frost_social_2008,levonian_patterns_2020} from others by receiving comments or reactions on their posts. However, health-specific platforms are predicated upon a specialized set of user needs, during what is often an extremely difficult and lengthy period of life~\cite{rains_social_2011,mcdaniel_new_2012,keating_health_2015,mccosker_living_2013}. Even though CMC affordances are similar in niche OHC platforms and general purpose OSNs, health-focused platforms must account for these underlying contextual differences, or they risk reducing their specialized value-proposition and benefits to users.

\UIappearance

In this study, we turn our focus toward CMC affordances for \textit{reacting} to sensitive health disclosures, and we conduct this work in collaboration with \url{CaringBridge.org}, a health-focused platform with over 45M users annually~\cite{caringbridge_about_2022}. Patient and caregiver users of CaringBridge have individual ``sites'' where they can post ``Journal'' updates about their health journey~\cite{ma_write_2017}. Since 2012, visitors to Journal updates have been able to write a comment and/or apply a \raisebox{-.4\mydepth}{\includegraphics[height=\myheight]{reactions/amp.jpg}} ``Heart'' reaction,\footnote{Internally to the organization (as well as in prior work~\cite{levonian_patterns_2020}), the symbol is referred to as ``amp,'' which is short for ``amplify'' and is represented by waves of care emitting from the top of the heart. However, we use the term ``Heart'', since that is how the symbol is labeled in CaringBridge's user interface.} which is the company logo at CaringBridge. This type of ``Liking'' affordance has become a ubiquitous feature in modern social media. In fact, Facebook’s Like button \raisebox{-.4\mydepth}{\includegraphics[height=\myheight]{emoji/fb_like.png}} is the most recognizable icon associated with the Social Web~\cite{gerlitz_like_2013}. Single-click liking features are Paralinguistic Digital Affordances (PDAs)~\cite{hayes_one_2016} that enable low-cost (\textit{i.e.}, quick) social engagement with meaningful social cues~\cite{ellison_cultivating_2014,scissors_whats_2016}. However, prior work raises a concerning question in the context of health-specific platforms~\cite{carr_as_2016,ahmadi_designing_2016}: Is this type of micro-interaction a meaningful expression of support, given the specialized needs of patients and caregivers?

Furthermore, several platforms (\textit{e.g.}, Facebook, Slack, LinkedIn, and others) have now transitioned from offering \textit{liking} more specifically to \textit{reacting} more broadly, providing users with a panel of emoji or sticker-like symbols to choose from. Facebook made this design decision in order to provide a more emotionally comprehensive affordance for quick reactive feedback, reflecting some of the most commonly-applied emoji and stickers in comments on posts~\cite{teehan_reactions_2016}. %
On January 27, 2021, following this trend in mainstream social media, CaringBridge also transitioned from the single Heart to a hover menu incorporating three additional reaction symbols: \raisebox{-.5\mydepth}{\includegraphics[height=\myheight]{reactions/pray.jpg}} ``Prayer''; \raisebox{-.5\mydepth}{\includegraphics[height=\myheight]{reactions/happy.jpg}} ``Happy''; and \raisebox{-.5\mydepth}{\includegraphics[height=\myheight]{reactions/sad.jpg}} ``Sad.'' (See Figure~\ref{fig:UI}.) This implementation thus combines a CaringBridge-branded Heart symbol, a support-specific Prayer symbol, and two emoji-like Happy and Sad face symbols that closely resemble Facebook's reactions. CaringBridge's launch of this feature provides a unique opportunity to investigate perceptions of each type of symbol for reacting to health-related posts, %
as well as how the transition from one to several reaction options affects the experience of support exchange on a health-specific platform. 

In order to better understand users' experiences of supportive communication in health-critical contexts, and to identify design opportunities to improve this experience, the present work examines the following research questions: %

\begin{description}
\item \textbf{RQ1:} How do users perceive the single-click Heart~\raisebox{-.5\mydepth}{\includegraphics[height=\myheight]{reactions/amp.jpg}} reaction in CaringBridge's health blogging context?
\item \textbf{RQ2:} How does expansion from the single-click Heart~\raisebox{-.4\mydepth}{\includegraphics[height=\myheight]{reactions/amp.jpg}} reaction into a larger set of four reactions, including Heart~\raisebox{-.5\mydepth}{\includegraphics[height=\myheight]{reactions/amp.jpg}}, Prayer~\raisebox{-.5\mydepth}{\includegraphics[height=\myheight]{reactions/pray.jpg}}, Happy~\raisebox{-.5\mydepth}{\includegraphics[height=\myheight]{reactions/happy.jpg}}, and Sad~\raisebox{-.5\mydepth}{\includegraphics[height=\myheight]{reactions/sad.jpg}}, impact users’ perceptions of reacting on CaringBridge? I.e. How do users perceive the...
\item \phantom{hello} \textbf{(a)} Emotion-based smileys (\raisebox{-.5\mydepth}{\includegraphics[height=\myheight]{reactions/happy.jpg}}, \raisebox{-.5\mydepth}{\includegraphics[height=\myheight]{reactions/sad.jpg}}) in this set?
\item \phantom{hello} \textbf{(b)} Non-face-based symbols (\raisebox{-.5\mydepth}{\includegraphics[height=\myheight]{reactions/amp.jpg}}, \raisebox{-.5\mydepth}{\includegraphics[height=\myheight]{reactions/pray.jpg}}) in this set?
\item \textbf{RQ3:} How does expanding CaringBridge's reactions feature affect broader dynamics of support exchange? I.e. How does this change impact...
\item \phantom{hello} \textbf{(a)} Authors’ needs and expectations for supportive communication?
\item \phantom{hello} \textbf{(b)} Visitors’ decision-making processes for expressing support?
\end{description}
We elicited the perceptions, expectations and opinions of 808 CaringBridge users via an online survey. %
We also conducted 13 follow-up interviews, thereby gaining richer perspectives on users' experiences of support exchange. Finally, we used Grounded Theory Method to analyze our qualitative data, and we computed descriptive statistics for quantitative survey questions.

We found that an expanded set of reactions transformed users' perceptions of the role of reactions in supportive communication. The primary contribution of this paper is an empirical description of how users of a health-focused blogging platform perceived the release of a new reactions feature. When there was only one Heart reaction, users believed it carried far more meaning than ``default liking'' on general purpose OSNs due to the high-stakes medical context of CaringBridge. Moreover, users typically interpreted Heart as an expression of simple acknowledgment (the update has been read) and support (the visitor truly cares). Although the majority of users \textit{accepted} the new reactions, our results also provide evidence that the product launch:

\begin{itemize}
    \item Increased the complexity of the interface and of supportive decision-making for users.
    \item Introduced \textit{emotional} ambiguity and confusion around how face-based expressions are supposed to function (\textit{e.g.,} self-expression? empathy? encouragement?).
    \item Introduced \textit{semantic} ambiguity about the meaning of non-faced based symbols (\textit{e.g.}, difficulty distinguishing the meaning of Heart \textit{v.s.} Prayer symbols).
    \item Raised concerns regarding which symbols/religions/racial identities are included or excluded.
\end{itemize}

Major disagreements and contention emerged around the meanings and functions of reactions in the high-stakes context of health crises. Some users found reactions useful, convenient, and reducing of caregiver burden; others believed they totally strip communication of meaningful expression and authentic care. In dialogue and resonance with recent work at CSCW discussing the importance of \textit{effortful} CMC~\cite{zhang_auggie_2022}, our work highlights crucial trade-offs between the cognitive effort, meaningfulness, and efficiency of different forms of CMC, and we illuminate design risks in this domain. In particular, improper symbolic reactions may harm authors during a vulnerable moment in their health journey, whereas overly complex PDA mechanisms may increase the burden of providing social support and promote more confusing communication in OHCs.
 
Overall, this study surfaces tensions for small social media platforms that need to survive amidst giants. Following mainstream trends may be expected by users and/or necessary for business, yet it may also threaten what feels most special and meaningful to users of more intimate online communities. In our discussion, we contribute: (1) design implications for improving health-focused CMC by reducing ambiguity, minimizing exclusion, and better aligning visitors' uses of reacting \textit{v.s.} commenting affordances with authors' CMC preferences and needs; and (2) a concrete set of research questions to guide studies of health-focused CMC in future work.

\section{Background Information on Online Health Communities}
\label{sec:background}

Because our primary contributions pertain to the use of reactions in OHCs, we begin by summarizing background information on OHCs across a variety of social media platforms (sec.~\ref{sec:OHCs}). Next, we provide more detail on our collaborative relationship with CaringBridge (sec.~\ref{sec:CB}). In the Related Literature (sec.~\ref{sec:RL}), we then describe prior work on emoji and reactions in CMC.

\subsection{Online Health Communities Across Social Media Platforms}\label{sec:OHCs}

Online health communities (OHCs) are diverse online spaces united by a focus on health~\cite{allison_logging_2021,gatos_how_2021}---including both stand-alone communities like CaringBridge~\cite{smith_i_2020}, PatientsLikeMe~\cite{frost_social_2008}, and the Cancer Survivors Network~\cite{yang_commitment_2017} and \textit{sub-}communities on larger OSNs like Facebook~\cite{young_this_2019}, Reddit~\cite{park_longitudinal_2017}, Instagram~\cite{andalibi_sensitive_2017}, or Tumblr~\cite{chancellor_recovery_2016}. OHCs have attracted attention in CSCW as spaces where \textit{social support} is exchanged~\cite{cutrona_controllability_1992,gatos_how_2021} for diverse health conditions (\textit{e.g.}, from cancer~\cite{allison_logging_2021} to mental health~\cite{sharma_mental_2018}). Social support has been theorized in many ways, but it generally includes both \textit{action-facilitating} forms of support---such as informational and instrumental assistance---as well as \textit{nurturant} forms, such as emotional, esteem, and network support~\cite{cutrona_controllability_1992}. Mirroring offline social norms, recent work at CSCW also demonstrates that \textit{spirituality} and the \textit{sense of community} derived from spiritual or religious organizations, groups, activities, and rituals often underlies and shapes the ways in which users exchange social support in OHCs~\cite{smith_i_2020,smith_what_2021,oleary_community_2022}. Access to social support has been linked to a variety of positive health and quality of life outcomes~\cite{rains_coping_2018,uchino_social_2004,holt-lunstad_social_2010}, and OHCs represent an opportunity to gain access to critical health-improving support~\cite{gatos_how_2021,allison_logging_2021}. Therefore, when users are managing chronic illness, they often participate across ``technology ecosystems'' in which they intentionally adopt, configure, and manage their behaviors across sets of platforms that help them to meet their complex health and privacy needs~\cite{burgess_technology_2020}. For example, on platforms such as PatientsLikeMe and Cancer Survivors Network where anonymous patients and caregivers interact with each other, informational support is highly prevalent~\cite{frost_social_2008,yang_commitment_2017}, whereas on platforms like CaringBridge~\cite{smith_i_2020} or medical crowdfunding platforms like GoFundMe~\cite{kim_designing_2022}, spiritual, emotional, and instrumental or financial support may be more common.

\subsubsection{Health Blogs}\label{sec:healthblogs}
Situated within users' health tech ecologies, health blogs are a particular subset of OHCs focused on individual patients and their ties to both existing (offline) support networks and new online connections with peers~\cite{rains_social_2011,mcdaniel_new_2012,keating_health_2015,mccosker_living_2013}. In contrast to casual ongoing use of ``general purpose'' OSNs, authors of health blogs often invest significant affective labor to describe how illness impacts their lives, and they derive important personal, social, and network-enabling value from these efforts~\cite{mccosker_living_2013}. The serious, life-threatening, and often long-term or chronic nature of these users' needs for supportive communication calls for specialized research attention in CMC and CSCW. For example, CaringBridge has been the focus of multiple CSCW studies (\textit{e.g.},~\cite{ma_write_2017,smith_i_2020,levonian_patterns_2020}) that describe the benefits of receiving supportive comments and reactions. CaringBridge blogs are most often followed by a network of known family, friends, co-workers, and members of the patient's offline communities (esp. spiritual or religious communities)~\cite{smith_what_2021}, however users occasionally seek out and form connections with strangers coping with similar health issues or roles~\cite{levonian_patterns_2020}. Because our work is situated on CaringBridge, we next provide background information about the platform and our collaborative partnership with them. 

\subsection{The CaringBridge Platform}\label{sec:CB}
As a 501(c)(3) non-profit organization, CaringBridge (\url{https://www.caringbridge.org}) provides free health blogging services, receiving traffic from nearly 400K visitors daily and 45M unique visitors total in 2021~\cite{caringbridge_about_2022}. Most CaringBridge blogs discuss cancer, but other conditions include premature births, surgeries, neurological disorders, and more~\cite{li_condition_2018,levonian_patterns_2020}. CaringBridge was founded in 1997, pre-dating today's mainstream OSNs, and has subsisted upon small donations from loyal users, grants, and large charitable endowments from individual donors. CaringBridge is a small organization with fewer than 100 employees. To highlight differing resource availabilities, we note that Facebook had over 70K employees~\cite{dixon_facebook_2022} and about 1.9B daily users in 2021~\cite{dixon_facebook_2022}. CaringBridge does not sell user data or use advertising to drive revenue, and users benefit from specialized privacy settings that enable authors to make their sites private and invite-only, publicly viewable by anyone, or viewable only by registered CaringBridge users. The last option is most popular because it enables authors to block unwanted visitors and monitor which visitors have seen which updates. Visitors can subscribe to Journals and receive email notifications when new updates are posted, enabling authors to inform their support communities of ongoing developments over the entire course of a health crisis. This study focuses primarily on affordances for responding to Journal updates (\textit{i.e.}, comments and reactions); the platform's complete set of technical features are defined in~\cite{ma_write_2017,smith_i_2020}. 

\subsubsection{The CaringBridge Research Collaborative}
Our work proceeds from a research collaboration formed in 2015 between CaringBridge and an interdisciplinary team of researchers originating from the University of Minnesota. The authors of this paper are \textit{not} employees of CaringBridge. However, we conducted this work collaboratively with CaringBridge, with the mutual aims of making important research contributions in CSCW, while also providing pragmatic insights to improve CaringBridge's products and ability to serve its users.
\section{Related Literature}\label{sec:RL}
In this section, we first summarize the growing body of work on emoji (sec.~\ref{sec:emojiCMC}), before focusing our lens upon emoji in their specific paralinguistic use as reactions (sec.~\ref{sec:reactionsCMC}). Finally, we highlight how reactions relate to the broader set of choices users make when exchanging support---\textit{i.e.}, reacting \textit{v.s.} commenting \textit{v.s.} responding offline (sec.~\ref{sec:supportCMC}).

\subsection{Emoji in Computer-Mediated Communication (CMC)}\label{sec:emojiCMC}
Emoji (derived from the Japanese for \textit{e} ``picture'' + \textit{moji} ``character'') are now a ubiquitous element permeating CMC. Emoji are single characters that sit in-line with text and are commonly used across CMC applications like messaging, email, and texting~\cite{dresner_functions_2010}. Billions of emoji are exchanged daily across platforms like Facebook, Messenger, Instagram, and Twitter~\cite{emojipedia_emoji_2022}. In 2015, about 2\% of Tweets~\cite{novak_sentiment_2015} and over 50\% of Instagram messages~\cite{dimson_emojineering_2015} already contained at least one emoji, with usage continuing to increase annually. For example, by 2018, 14.3\% of Tweets contained at least one emoji, rising to 15.7\% in 2019, 18.7\% in 2020, and 20.1\% in 2021~\cite{broni_top_2021}. In 2021, the following 10 emoji were used most commonly worldwide: \raisebox{-.5\mydepth}{\includegraphics[height=\myheight]{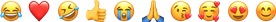}}, with smileys consistently dominating the list, and \raisebox{-.5\mydepth}{\includegraphics[height=\myheight]{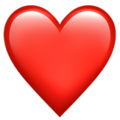}} and \raisebox{-.5\mydepth}{\includegraphics[height=\myheight]{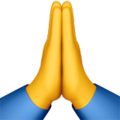}}, ranking 2nd and 6th respectively not only in 2021, but also several preceding years~\cite{daniel_most_2021}.

\subsubsection{What major functions in CMC do emoji serve?} In a 2019 systematic literature review, \citeauthor{bai_systematic_2019} demonstrate the expanding breadth of emoji research across scientific disciplines~\cite{bai_systematic_2019}. Querying the term ``emoji'' on Google Scholar and Web of Science, the authors survey 167 papers after 1998, primarily in the fields of Computer Science (30.18\%) and Communication (26.04\%), as well as Marketing (14.79\%), Behavioral Science (12.43\%), Linguistics (5.92\%), Psychology (5.33\%), Medicine (2.96\%), and Education (2.37\%). While these works examine a variety of contexts and applications of emoji use, there is some consensus that compared to text and emoticons\footnote{Emoticons (``emotion'' + ``icon'') emerged in the 1980s, and are sequences of characters that mimic the appearance of an emotional face--\textit{e.g.}, ;-D)~\cite{lang_electronic_2009,jibril_relevance_2013}}, emoji are perceived as richer visual representations that are: easier and more convenient to input; more technically advanced; and able to convey meaning through alternate sensory processing modes~\cite{aull_study_2019,barbieri_how_2016,ai_untangling_2017}. Thus, emoji help to fill cognitive gaps left in the transition from multi-sensory, face-to-face interaction to more austere online or text-based interactions~\cite{gantiva_cortical_2020}. 

\citeauthor{bai_systematic_2019} highlight how emoji can serve two major functions in CMC: First, emoji can provide non-verbal cues for expressing emotion or intention~\cite{alshenqeeti_are_2016,riordan_communicative_2017,riordan_emojis_2017}, thereby helping to maintain interpersonal relationships and manage conversational tone online~\cite{kelly_characterising_2015,chairunnisa_analysis_2017,albawardi_translingual_2018}. Second, emoji can also express other semantic meaning(s) beyond emotions~\cite{naaman_varying_2017}, even independently of words, as a form of visual rhetoric~\cite{jibril_relevance_2013}. These meanings are highly context-dependent~\cite{gawne_emoji_2019}, which leads to ambiguity in how emoji are supposed to be used when they appear in different settings~\cite{jaeger_valence_2019}.

\subsubsection{How do people use and understand emoji?}
Research has demonstrated high diversity in emoji adoption and usage across individual characteristics (\textit{e.g.}, generation~\cite{alshenqeeti_are_2016}, gender~\cite{herring_receiver_2018,prada_motives_2018,tossell_longitudinal_2012,chen_through_2018,rodrigues_lisbon_2017}, emotional wellness~\cite{settanni_sharing_2015}, extraversion~\cite{hall_self-monitoring_2013,li_mining_2018}), cultural/geographic contexts (\textit{e.g.}, Japanese teenagers~\cite{sugiyama_kawaii_2015}, Finland/India/Pakistan~\cite{sadiq_learning_2019}, China/Spain~\cite{cheng_i_2017}, Hong Kong/USA~\cite{chik_comparative_2017}), and platforms (\textit{e.g.}, IOS vs. Android~\cite{rodrigues_lisbon_2017,miller_blissfully_2016,miller_hillberg_what_2018} or Facebook vs. Twitter vs. Instagram~\cite{tauch_roles_2016}). Thus, emoji usage is highly contingent upon technical, linguistic and textual environments~\cite{vandergriff_emotive_2013}.

Research has also demonstrated extensive variance in how people interpret emoji, both in terms of semantic meaning and sentimental valence~\cite{miller_blissfully_2016,tigwell_oh_2016}. In the Unicode Consortium, each emoji is designated by a unique code (or combination of codes), however platforms render these codes differently (\textit{e.g.}, ``woman'' (U+1F469) is rendered \raisebox{-.5\mydepth}{\includegraphics[height=\myheight]{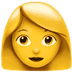}} by Apple; \raisebox{-.5\mydepth}{\includegraphics[height=\myheight]{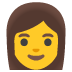}} by Google; and \raisebox{-.5\mydepth}{\includegraphics[height=\myheight]{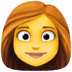}} by Facebook~\cite{unicode_full_2022}). The possibility for misaligned interpretations of emoji are worsened by divergent cross-platform renderings~\cite{miller_blissfully_2016}. Many users are unaware of these rendering differences, and say they would change past behaviors (\textit{e.g.}, editing or not sending a Tweet), had they known~\cite{miller_hillberg_what_2018}. Moreover, emoji are more frequently set within surrounding text than used in isolation~\cite{medlock_linguistic_2016,zhou_goodbye_2017}. Some work suggests that this textual embedding can clarify the intention, meaning, and credibility of emoji~\cite{alshenqeeti_are_2016,zhou_goodbye_2017,daniel_emojis_2018}, whereas other work demonstrates that misinterpretation of emoji is not reliably improved by surrounding text~\cite{miller_understanding_2017}.

\subsubsection{How are emoji used in business and marketing?}
Because \raisebox{-.5\mydepth}{\includegraphics[height=\myheight]{reactions/amp.jpg}} is CaringBridge's logo, we will also contextualize what is known about business uses of emoji. Brand logo benefit has been demonstrated to help improve commitment of stakeholders to organizations, such as universities~\cite{japutra_whats_2016}. To our knowledge, no prior work has evaluated the use of a branded logo as a reaction.\footnote{Facebook's ``Like'' \raisebox{-.4\mydepth}{\includegraphics[height=\myheight]{emoji/fb_like.png}} is not the company logo, although it is clearly brand-affiliated.} Marketing research suggests that emoji do impact consumer decisions and behaviors, such as willingness to purchase a product~\cite{das_emoji_2019} or engage with advertising messages~\cite{luangrath_textual_2017,manganari_enhancing_2017}. For example, social media influencers often initiate messages with emoji in order to attract consumers~\cite{ge_emoji_2018}; this may be particularly effective for engaging younger audiences~\cite{yakin_application_2017}. Since businesses use emoji to understand consumers' emotional reflections on their products, services, or branding~\cite{moreno-sandoval_spanish_2018,rathan_every_2017,moussa_emoji-based_2019}, some work also explores evaluative uses of emoji-based surveys. For example, \citeauthor{jaeger_emoji_2018} studied a survey tool based on 33 facial emoji, finding that consumers are generally able to interpret and discriminate between these emoji, with very few differences based on age, gender, or frequency of emoji usage~\cite{jaeger_emoji_2018}; yet such surveys should not rely \textit{exclusively} on emoji, but should consider them as complementary to text-based ways of understanding sentiment~\cite{jaeger_measurement_2017,jaeger_measuring_2018}. Similarly, \citeauthor{kaur_i_2022} also found that emoji-based surveys provide some degree of signal about workers' wellbeing on the job, but cannot replace text or other measurement inputs~\cite{kaur_i_2022}.

\subsection{Reactions as Paralinguistic Digital Affordances}\label{sec:reactionsCMC}
In this section, we summarize the history and prior work on reactions, beginning with some of their initial appearances across prominent social media platforms, and then highlighting how interfaces for reacting have evolved over time. We characterize a variety of \textbf{reaction-styles}, using boldfaced font to indicate terms that we will use throughout the paper, and we highlight the gaps in prior work that have inspired our research questions.

Reactions have a similar (or identical) appearance to emoji. Rather than sitting in-line with text, however, users apply reactions to complete posts or messages written by others. Reactions are characterized as Paralinguistic Digital Affordances (PDAs) because they are \textit{``cues in social media that facilitate communication and interaction without specific language associated with their messages.''}~\cite{hayes_one_2016} Moreover, in in-person contexts, \textit{phatic} communication includes words or phrases that \textit{``fulfill a social function ... but they are neither the result of intellectual reflection, nor do they necessarily arouse reflection in the listener''}~\cite{malinowski_phatic_1972}. For example, people customarily initiate conversations with exchanges like, \textit{``How are you?''} and \textit{``I'm good''}; this exchange contains limited meaningful information but fulfills an expected function. In online communication, PDAs have been theorized as a phatic form of communication that allows people to adopt a minimalist communication strategy~\cite{coupland_how_1992}, since some exchanges on social media lack substantive content~\cite{miller_new_2008}.

\subsubsection{The origin of reactions as a single-option, single-click affordance}
Reactions were first implemented as a \textbf{``single-option, single-click''} affordance. As each new ``reactor'' clicks, the count of reactions is incremented, with that users' name included in a list of all reactors. For example, the ``Like'' \raisebox{-.4\mydepth}{\includegraphics[height=\myheight]{emoji/fb_like.png}} was introduced as Facebook's first reaction in 2009, and has since become one of the most recognizable icons across social media~\cite{gerlitz_like_2013}. Similarly, Twitter's star-shaped ``Favorite'' button \raisebox{-.4\mydepth}{\includegraphics[height=\myheight]{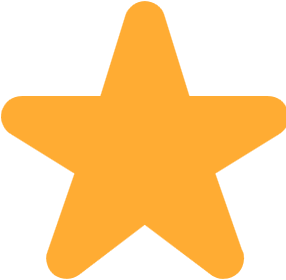}} was implemented in 2006 (pre-dating Facebook's ``Like''), but was later transformed and rebranded in 2015 into a heart-shaped ``Like'' \raisebox{-.4\mydepth}{\includegraphics[height=\myheight]{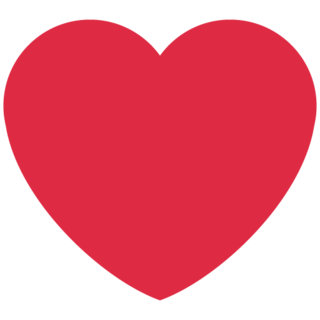}}~\cite{kumar_hearts_2015}. Instagram added its similar heart-shaped ``Like'' \raisebox{-.4\mydepth}{\includegraphics[height=\myheight]{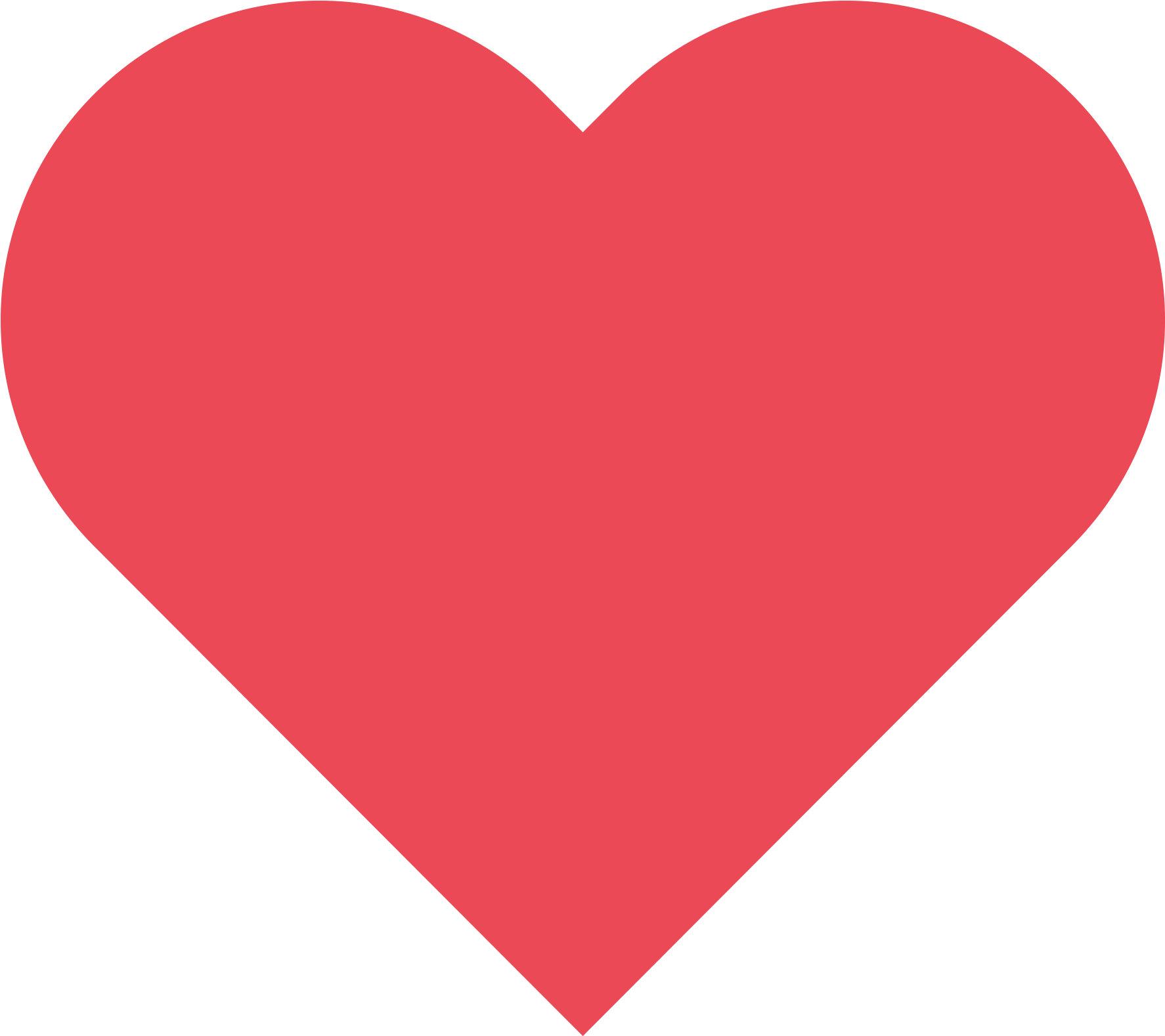}} in 2015~\cite{hallinan_living_2021}, whereas CaringBridge added its ``Heart'' \raisebox{-.4\mydepth}{\includegraphics[height=\myheight]{reactions/amp.jpg}} logo as a reaction in 2012. These examples are not exhaustive, yet they suggest some diversity in company's selection of reactions, as well as convergence toward heart-shaped single-click PDAs.

As with emoji, prior work demonstrates a tremendous amount of diversity in people's motivations, uses, and gratifications of single-option, single-click reactions~\cite{hayes_one_2016}, with a predominant focus on ``Like'' buttons. On Facebook, Likes are described as a ``social button''~\cite{gerlitz_like_2013} that can be used for lightweight relationship maintenance and interpersonal communication~\cite{ellison_cultivating_2014,scissors_whats_2016,marwick_status_2013}, social bookmarking~\cite{heymann_can_2008}, social affirmation and sense of belonging~\cite{scissors_whats_2016,reich_zero_2018}, virtual endorsement~\cite{lee_effect_2014}, affective processing~\cite{gehl_reverse_2014}, or automated feedback~\cite{hearn_structuring_2010}. On Instagram, artists use the Like button for overlapping affective, relational, and infrastructural evaluations~\cite{hallinan_living_2021}.

Although most studies examine Likes in a generalized sense, some papers have examined the degree to which single-click PDAs can be perceived as social support. A 2016 survey of users of five popular social media ($N=325$, mostly Facebook (64.6\%), with less than 10\% each across Twitter, Pinterest, Instagram, and LinkedIn) found three factors impacting perceived supportiveness: (1) whether or not their post was seeking support; (2) relational closeness with the PDA sender; and (3) how \textit{automatic} the receiver perceives the PDA to be~\cite{carr_as_2016}. People also seek different aspects of support on different platforms, depending on both the platform itself and the audiences it serves~\cite{hayes_its_2016}. For example, focus groups with college students suggest that Likes feel inappropriate for situations when friends have posted about negative experiences (\textit{e.g.}, stressful courses, break-ups, etc.); to ``Like'' the negativity does not feel like a real expression of support, but something like ``Hug'' might be better~\cite{ahmadi_designing_2016}. A different study of college students suggested that even if something difficult is disclosed on Facebook (\textit{e.g.}, failing an exam), its author may nonetheless interpret Likes as support~\cite{hayes_one_2016}. Another study used an ``empathy button'' (with a similar appearance to ``Like'') within an app designed to raise communal awareness of trending stress-related topics among college students. Results demonstrated that this empathic feature helped to foster \textit{``reactive resonance''} and to provide timely, anonymous, and compassionate peer support~\cite{choi_you_2022}.

From these works and user populations, we know that single-click Likes \textit{can} be perceived as supportive in relation to negative experiences, yet significant tensions or barriers exist related to their usage on general-purpose social media, where they can mean many different things. We do not know how people perceive single-click PDAs in the narrower context of health- and social support-specific platforms. The present work thus contributes ecologically-derived knowledge of users' perceptions of a single-option, single-click reaction on CaringBridge--an impactful real world health-support platform--thus addressing this gap. Our first research question is:

\begin{description}
\item \textbf{RQ1:} How do users perceive the single-option, single-click \raisebox{-.4\mydepth}{\includegraphics[height=\myheight]{reactions/amp.jpg}} reaction in CaringBridge's  health blogging context?
\end{description}

\subsubsection{The evolution of reactions into multi-option, multi-click affordances}
Over time, the technical affordances of reacting have evolved across some platforms to involve multiple options and series of clicks. For example, Slack reactions were launched in 2015 with a more complex reaction-style that we refer to as a \textbf{``free-for-all.''} Similar to selecting emoji from a menu, Slack enabled users to create a reaction out of \textit{any} emoji supported by Slack~\cite{mccracken_slacks_2015}, including unofficial ``slackmojis'' that are custom-designed by users~\cite{kupferman_slackmojis_nodate}. After a first reaction has been added, subsequent reactors can either click on an existing reaction (visibly incrementing its count), and/or add a different reaction, such that a post can have many different types of reactions applied.

In contrast to the free-for-all, we refer to the reaction-style adopted by CaringBridge as a \textbf{``limited-bar.''} (See Fig.~\ref{fig:UI}.) Facebook popularized this style in February 2016, when it expanded the single-click Like into a restricted set of options, including Like~\raisebox{-.5\mydepth}{\includegraphics[height=\myheight]{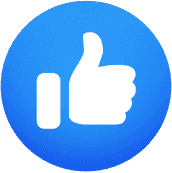}}, and new custom-designed animations of Love~\raisebox{-.5\mydepth}{\includegraphics[height=\myheight]{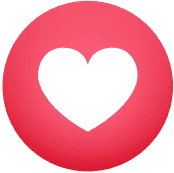}}, Haha~\raisebox{-.5\mydepth}{\includegraphics[height=\myheight]{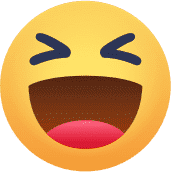}}, Wow~\raisebox{-.5\mydepth}{\includegraphics[height=\myheight]{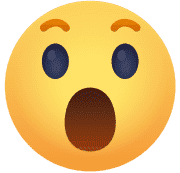}}, Sad~\raisebox{-.5\mydepth}{\includegraphics[height=\myheight]{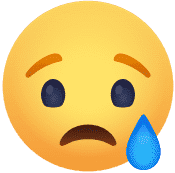}}, and Angry~\raisebox{-.5\mydepth}{\includegraphics[height=\myheight]{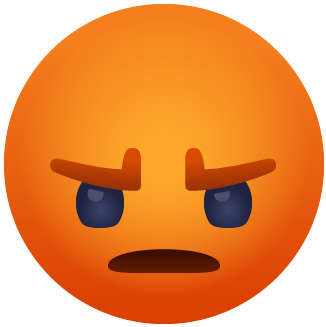}}~\cite{teehan_reactions_2016}, later adding Care~\raisebox{-.5\mydepth}{\includegraphics[height=\myheight]{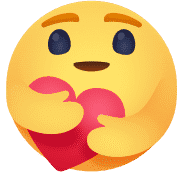}} in 2020 as a response to the COVID-19 pandemic~\cite{facebook_facebook_2020}. (Interestingly, Facebook also initially experimented with \textit{temporary} reactions (\textit{e.g.}, Thankful~\raisebox{-.5\mydepth}{\includegraphics[height=\myheight]{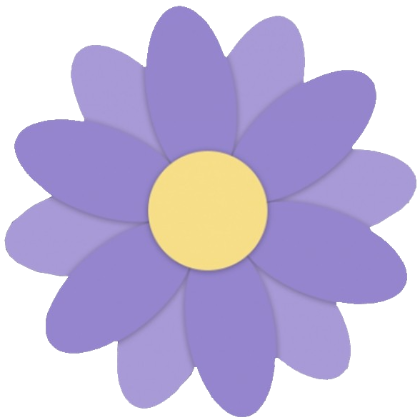}} and Pride~\raisebox{-.5\mydepth}{\includegraphics[height=\myheight]{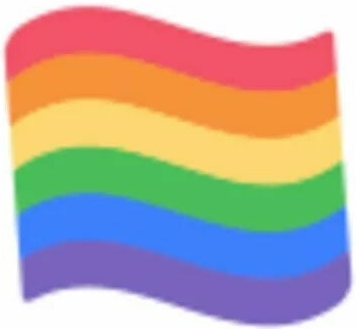}} in 2016-17) before announcing their permanent discontinuation~\cite{liptak_facebook_2018}.) As a point of comparison, LinkedIn launched a limited-bar of reactions in 2019, including Like~\raisebox{-.5\mydepth}{\includegraphics[height=\myheight]{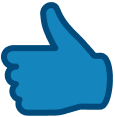}}, Celebrate~\raisebox{-.5\mydepth}{\includegraphics[height=\myheight]{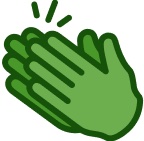}}, Support~\raisebox{-.5\mydepth}{\includegraphics[height=\myheight]{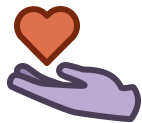}}, Love~\raisebox{-.5\mydepth}{\includegraphics[height=\myheight]{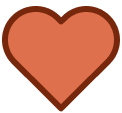}}, Insightful~\raisebox{-.5\mydepth}{\includegraphics[height=\myheight]{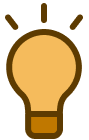}}, and Curious~\raisebox{-.5\mydepth}{\includegraphics[height=\myheight]{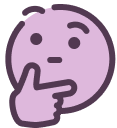}}~\cite{chen_introducing_2019}, later adding Funny~\raisebox{-.5\mydepth}{\includegraphics[height=\myheight]{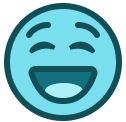}} in 2022~\cite{ahmed_linkedin_2022}. We draw attention to design choices such as skin tones, as well as similarities and divergences in these platforms' chosen sets of smiley and symbolic or gesture-based graphics; we will explore these types of design decisions in our discussion. 

Our work primarily considers single-option, single-click \textit{v.s.} the free-for-all \textit{vs.} the limited-bar. However, we acknowledge that other important forms of reacting and paralinguistic PDAs exist on platforms with up/downvoting features\footnote{One notable omission is Reddit's ``Upvote'' \raisebox{-.4\mydepth}{\includegraphics[height=\myheight]{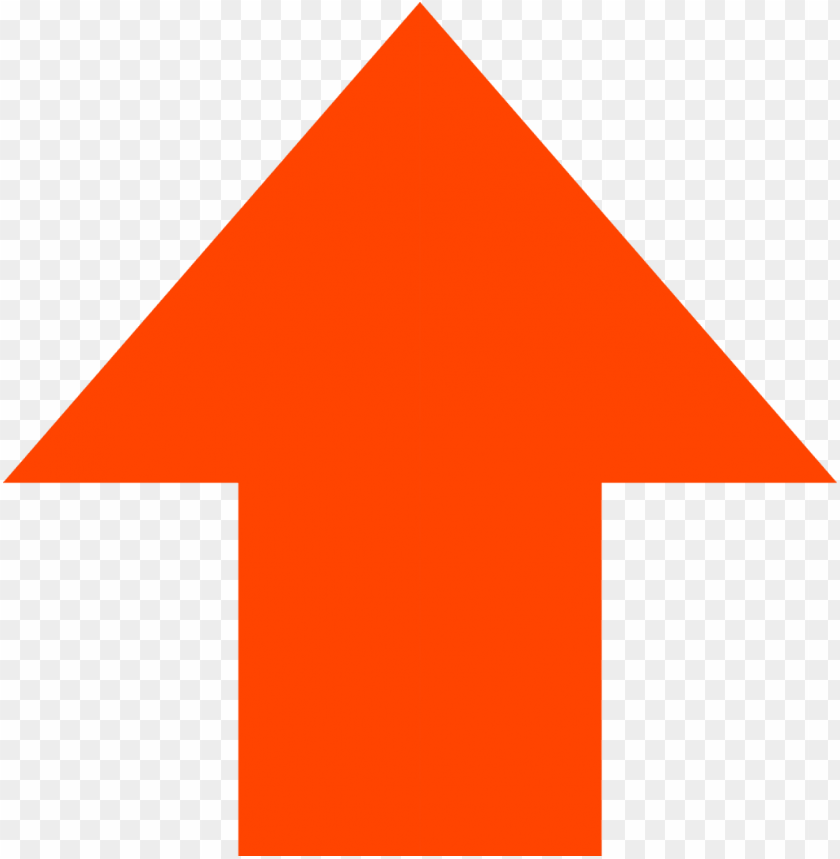}} and ``Downvote'' \raisebox{-.4\mydepth}{\includegraphics[height=\myheight]{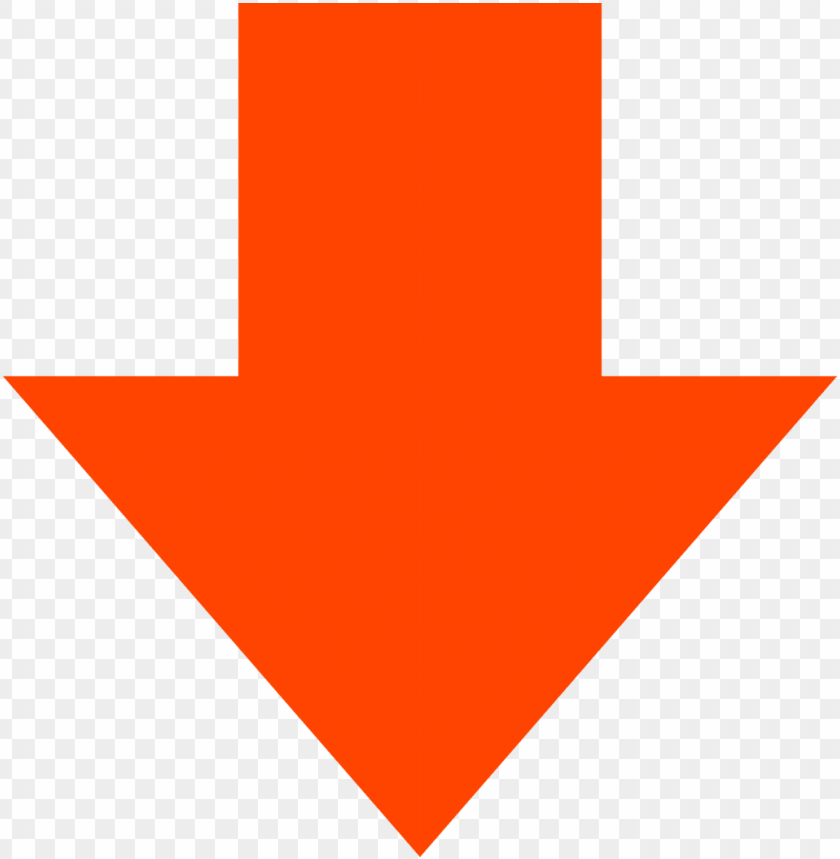}} feature, which has major design divergences from the reaction-styles we focus on in the present work---\textit{i.e.}, the count of up/down votes displayed is: an aggregation of up- minus downvotes, anonymous, and \textit{transparently} impacts ordering of content and users' ``karma'' scores. Subreddit moderators can also customize the up- and downvote functions with any symbols of their choice.} or real-time video streaming, video meetings or webinars, and mobile text messaging, etc. that we do not cover here.

\subsubsection{The launch of emoji reactions on Facebook}
\citeauthor{wisniewski_happiness_2020} studied the launch of Facebook Reactions in 2016 by collecting and analyzing users' comments on two public announcements posted by Facebook CEO, Mark Zuckerberg~\cite{wisniewski_happiness_2020}. The first post described how and why the Like feature would soon be expanded into a set of emotion-based choices, while the second marked the public launch date of the new feature. The authors discuss users' desires and apprehensions surrounding imagined uses of a controversial Dislike button (which was never actually released on Facebook, although it now exists in other CMC contexts). However, they find that overall, after using the new feature, users appreciated the affective capabilities of reactions, even though they cannot express the full spectrum of emotion. Moreover, they noted users' desires for even more reactions--with one suggestion pointing toward users' interest in a \raisebox{-.5\mydepth}{\includegraphics[height=\myheight]{emoji/pray.png}} option. In our work, we have the opportunity to evaluate users' perceptions of the real-world launch of CaringBridge's version of this graphic, \raisebox{-.5\mydepth}{\includegraphics[height=\myheight]{reactions/pray.jpg}}, which they label Prayer.

\subsubsection{Prayer and spiritual support in CSCW}
The Unicode Consortium designates \raisebox{-.5\mydepth}{\includegraphics[height=\myheight]{emoji/pray.png}} by code U+1F64F with the name ``folded hands''~\cite{unicode_full_2022}, while Emojipedia details diverse possible meanings of the symbol, including prayer hands, please or thank you, namaste, greetings, or even high five~\cite{emojipedia_person_2022}. Despite the many meanings this symbol carries, it clearly holds major value to emoji-users worldwide, given its consistently high frequency of use~\cite{daniel_most_2021}. We posit that many of these usages likely carry spiritual connotations--especially for users in need of support~\cite{smith_sacred_2022}.

Although relatively little research in HCI has examined religion and spirituality~\cite{bell_no_2006,bell_chi_2010,buie_spirituality_2013,ahmad_emergence_2013,buie_let_2019}, prior work on CaringBridge has surfaced the deeply meaningful role of prayer~\cite{smith_i_2020} and spiritual support~\cite{smith_what_2021} in users' health-related CMC on the platform. %
Behaviorally, CaringBridge authors write more expressions of appreciation for prayer than any other form of support. Moreover, when surveyed, both authors and visitors rate prayer as the most important form of support exchange~\cite{smith_i_2020}. Peoples' underlying spiritual or religious beliefs often motivate their goals and choices for which forms of support they want to offer and receive--and consequently, how they want technology to support these aims~\cite{smith_what_2021}. For example, \cite{smith_what_2021} suggests design implications that improving techniques for visualization of spiritual support may deepen and improve peoples' feelings of connection and wellness. Examples include:
\begin{itemize}
    \item UI/UX elements and symbols that share one's beliefs (\textit{e.g.}, emoji, stickers, profile badges).
    \item Interactive visualizations showing the scope, scale, and capabilities of spiritual care networks.
\end{itemize}

Recent work also highlights technospiritual affordances for prayer in other contexts. For example, \citeauthor{kaur_sway_2021} built an interactive visualization tool representing spiritual support networks as gardens of flowers, with petals that represent prayer~\cite{kaur_sway_2021}. \citeauthor{stowell_investigating_2020} and \citeauthor{oleary_community_2022} built an application in partnership with Black churches that offers a prayer wall, along with other health-promoting features~\cite{stowell_investigating_2020,oleary_community_2022}. \citeauthor{lambton-howard_blending_2021} collaborated with three churches in Barbados in order to re-appropriate WhatsApp for the formation of peer support groups to promote weight loss for diabetes management. The purpose was to enable peer support to seamlessly ``blend into everyday life'' by connecting into existing church communities~\cite{lambton-howard_blending_2021}. Driven by the COVID-19 pandemic, recent work also explores the increasing prevalence of online Christian worship ceremonies~\cite{wolf_spirituality_2022} and Buddhist rituals such as communal chanting~\cite{claisse_keeping_2023}. Tying in with this growing body of work on spiritual and/or religious prayers and rituals, CaringBridge's Prayer~\raisebox{-.5\mydepth}{\includegraphics[height=\myheight]{reactions/pray.jpg}} reaction offers a compelling research avenue for furthering our knowledge of affordances for spiritual support.

\subsubsection{Understanding sets of reactions}
Facebook published its rationale and design process for selecting a set of emotionally expressive reactions in a Medium post~\cite{teehan_reactions_2016}. Users generally approved of the launch since it allows them to easily express more emotions~\cite{wisniewski_happiness_2020}. Does this concordance in design intent and users' reception of a new reaction-style generalize across platforms? Is expanding a single-option, single-click reaction-style into a limited-bar \textit{always} a good idea? And how might context-specific factors influence users' perceptions of the specific \textit{sets} of symbols chosen by platforms? CaringBridge did not publicly publish its design process, however our results summarize their organizational rationale for selecting this set of reactions (Sec.~\ref{sec:rationale}), which now includes two smileys and two symbols. Offering a point of rich juxtaposition against other OSNs, we ask:

\begin{description}
\item \textbf{RQ2:} How does expansion from the single-click Heart~\raisebox{-.4\mydepth}{\includegraphics[height=\myheight]{reactions/amp.jpg}} reaction into a larger set of four reactions, including Heart~\raisebox{-.5\mydepth}{\includegraphics[height=\myheight]{reactions/amp.jpg}}, Prayer~\raisebox{-.5\mydepth}{\includegraphics[height=\myheight]{reactions/pray.jpg}}, Happy~\raisebox{-.5\mydepth}{\includegraphics[height=\myheight]{reactions/happy.jpg}}, and Sad~\raisebox{-.5\mydepth}{\includegraphics[height=\myheight]{reactions/sad.jpg}}, impact users’ perceptions of reacting on CaringBridge? I.e. How do users perceive the...
\item \phantom{hello} \textbf{(a)} Emotion-based smileys (\raisebox{-.5\mydepth}{\includegraphics[height=\myheight]{reactions/happy.jpg}}, \raisebox{-.5\mydepth}{\includegraphics[height=\myheight]{reactions/sad.jpg}}) in this set?
\item \phantom{hello} \textbf{(b)} Non-face-based symbols (\raisebox{-.5\mydepth}{\includegraphics[height=\myheight]{reactions/amp.jpg}}, \raisebox{-.5\mydepth}{\includegraphics[height=\myheight]{reactions/pray.jpg}}) in this set?
\end{description}

\subsection{Beyond reactions alone: Frameworks for supportive communication and decision-making in CMC}\label{sec:supportCMC}
Thus far, our review of the literature has focused on emoji and reactions. However, an entire subarea of communications focuses on supportive communication, suggesting that \textit{``the study of social support is the study of supportive communication: verbal and nonverbal behaviors intended to provide or seek help.''}~\cite{macgeorge_chapter_2011} It is crucial to note that PDAs such as reactions represent only one possible choice of nonverbal support mechanism, situated within much broader ecologies of CMC technologies and response types in healthcare~\cite{burgess_technology_2020,ongwere_challenges_2022}. Additional verbal responses could include written comments online, cards sent in the mail, or phone calls, whereas additional non-verbal responses could include gestures like sending flowers or gifts, hugs, gentle touch, silent prayers, and dropping off a lasagna (or three). Two important questions arise: First, what is the interplay like between these different types of responses (or combinations of responses)~\cite{wan_how_2020,wisniewski_happiness_2020}? Second, how well-aligned are visitors' responses to authors' needs and preferences~\cite{smith_i_2020,levonian_patterns_2020,andalibi_responding_2018,andalibi_social_2018}? In this paper, we focus primarily on two possible responses (reactions and comments), and the alignment between visitors' and authors' communication preferences, later discussing directions for future work to contribute further to these broad questions.

\subsubsection{Reactions v.s. comments}
\citeauthor{wisniewski_happiness_2020} raise the possibility that Facebook's reaction launch might impact users' commenting behaviors~\cite{wisniewski_happiness_2020}. Since the launch, some studies have used reactions as a proxy metric for sentiment analysis, showing that when certain types of reactions are applied, they may change the frequency of other user behaviors, like sharing or commenting~\cite{kim_like_2017,smoliarova_emotional_2018,tran_sentiment_2018}, or provide informative cues about users' perceptions of technology (\textit{e.g.}, application of the ``Laugh'' reaction to comments made by bots on GitHub~\cite{farah_exploratory_2022}). Thus, increasing reactions can have important downstream effects on users. On CaringBridge, we know that sites which do not receive \textit{any} interaction from visitors have almost no chance of continuing past the first 1 or 2 Journal updates~\cite{wan_how_2020}. However, receiving supportive comments from visitors contributes to authors' sustained use of the platform~\cite{ma_write_2017}--a result that is often replicated across studies of online health communities, \textit{e.g.},~\cite{wang_stay_2012,wang_social_2014,yang_commitment_2017}. Prior to the launch of the new reaction options, CaringBridge sites that receive \textit{both} comments and Hearts are more likely to survive than those which receive exclusively comments \textit{or} Hearts alone, and receiving \textit{``many comments is more beneficial for survival than many [Hearts], but in small amounts receiving [Hearts] is more important than receiving comments.''}~\cite{wan_how_2020} Increasing the number of reaction options is likely to \textit{diversify} reacting behaviors (potentially increasing or decreasing total number of reactions applied), and may impact commenting behaviors (potentially increasing or decreasing the quantity and quality of comments)---thus, we expect that expanding reactions will have important repercussions for user engagement on CaringBridge.

\subsubsection{Understanding how people choose supportive responses}
Across platforms like Facebook, Instagram, and Reddit, Andalibi and her colleagues have developed frameworks for understanding why users may choose to engage in sensitive disclosures about health issues~\cite{andalibi_understanding_2016}, how viewers decide to respond (or not respond) to such disclosures~\cite{andalibi_responding_2018}, and whether these choices seem to align~\cite{andalibi_social_2018,andalibi_sensitive_2017}. Inspired by this approach, and by the knowledge that CaringBridge authors and visitors have previously been shown to diverge in at least \textit{some} of their needs and preferences for support exchange (\textit{e.g.}, instrumental support~\cite{smith_i_2020}) and patterns of connection with other users~\cite{levonian_patterns_2020}, we ask:

\begin{description}
\item \textbf{RQ3:} How does expanding CaringBridge's reactions feature affect broader dynamics of support exchange? I.e. How does this change impact...
\item \phantom{hello} \textbf{(a)} Authors’ needs and expectations for supportive communication?
\item \phantom{hello} \textbf{(b)} Visitors’ decision-making processes for expressing support?
\end{description}

\section{Methods}
Prior to designing our methods, we recorded a Zoom consultation with three CaringBridge employees in order to understand the motivation for releasing new reactions on the platform. (Prior work has similarly engaged with internal employee stakeholders (\textit{e.g.,} Stack Overflow~\cite{mamykina_design_2011}, Wikipedia~\cite{smith_keeping_2020}, etc.) Informed by this consult, we then selected two methods--surveys and interviews--to gather user perceptions of the launch of new reactions. Our survey allowed us to gather opinions and impressions from a large, representative sample of CaringBridge users. Next, we enriched and elaborated upon findings from the survey by conducting interviews with a smaller sample of survey participants who opted in to be contacted for an interview. 

\begin{figure}[b]
    \centering
    \includegraphics[width=\textwidth]{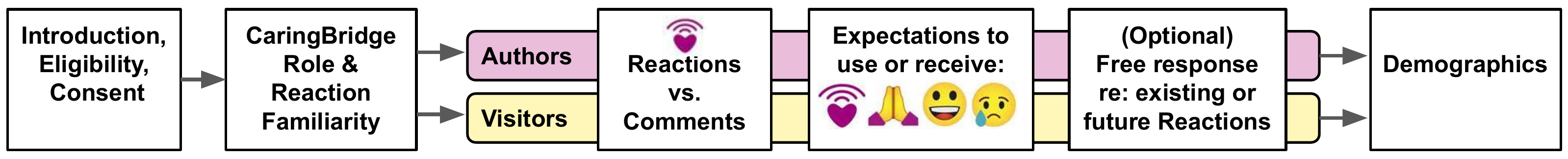}
    \caption{Survey Structure. Survey was built and deployed using Qualtrics software~\cite{qualtrics_qualtrics_2021}.}
    \label{fig:survey_flow}
\end{figure}

\subsection{Survey}
\subsubsection{Survey Design}
Figure~\ref{fig:survey_flow} shows an outline of the survey design. Throughout the survey, we used reaction images rather than words in order to avoid biasing users' interpretations of these images. The survey begins with a description of purpose, terms, and consent/eligibility questions. Next, users select their role on CaringBridge (\textit{i.e.}, author, visitor), and their role in current health journey(s) (\textit{i.e.}, self (patient), family caregiver, close family/friend, or casual acquaintance). The survey asks two similar sets of questions for authors and visitors eliciting opinions about: the importance of reactions vs. comments; usage expectations for reactions; any other thoughts on individual reactions or how CaringBridge should change its reactions. Finally, we asked for limited demographics (gender, age, spiritual beliefs), and whether users opt-in to share their CaringBridge email; participate in a followup interview; or receive an email about the study results. (See supplemental materials for verbatim survey text.)

\participants

\subsubsection{Survey Deployment}
An important methodological consideration for our study was whether to deploy the survey before, on, or after the launch date of January 27, 2021; this choice impacts how many users may have already used or been exposed to the new reactions feature. From one perspective, the choice is somewhat arbitrary, given that most users have already been exposed to the limited-bar reaction style on other social media. They understand how reacting works in general and are unavoidably biased by prior experiences of reacting in other contexts. Since we displayed the new reaction symbols within the survey, we view participants as well-equipped to express their initial perceptions of how the limited-bar reactions feature translates to the CaringBridge context, regardless of whether or not they had already used the feature. Yet actual usage of a new feature can also alter people's opinions. With our unique opportunity to study the \textit{transition} from one reaction to multiple, we decided that our research questions were best informed by a mix of perspectives, both from users who have and have not used the new feature yet. Qualitatively, participants who have can contribute opinions about their initial experiences, whereas users who have not (or who will deliberately choose not to) can explain their preferences and decisions. Quantitatively, it is possible to separate out users who have and have not yet used the feature and to discern whether or not significant differences exist in their numerical survey responses.\footnote{We performed a subgroup analysis to compare numerical responses from users who had and had not used the new reactions. Because we found no significant differences that would impact our main results, we present quantitative summaries for all respondents together. We describe the subgroup analysis and observed differences in Appendix~\ref{sec:subgroups}.}

Given all of these considerations, we intentionally chose a timeframe shortly after the launch and captured a broad range of experiences from users who had and had not used the new reactions feature. In order to recruit participants, CaringBridge posted a banner announcement inviting both authors and visitors to share their thoughts about communicating with reactions by taking our survey. This banner was posted from February 8--28, 2021 (excluding February 13--15). Only registered users have the ability to use reactions on CaringBridge; thus any registered CaringBridge users (ages 18+) were eligible. Participation was opt-in, voluntary, and without compensation.

\subsubsection{Survey Participants}
Table~\ref{tab:participants}a provides a summary of all respondents. Participant IDs beginning with an ``S'' indicate ``Survey Respondent.'' We collected 808 complete responses,\footnote{256 additional respondents were eliminated by eligibility criteria, while 138 eligible respondents who began the survey did not complete it, thus their responses were not included.} aiming to recruit a representative sample size similar to~\cite{smith_i_2020}. 437 respondents indicated their role as author of a CaringBridge site; the remaining 371 indicated visitor. Consistent with prior CaringBridge surveys~\cite{smith_i_2020}, the majority of respondents are female (85.5\%). Most are middle-aged, with 23.0\% 45-64 years old, and 27.8\% 55-64; an additional 27.2\% selected age buckets younger than 44, and 21.9\% 65 or older. Most participants are Christian (67.2\%) or Spiritual/Non-Religious (14.6\%); the remaining 18.2\% specified Jewish, Muslim, Atheist, Agnostic, or other beliefs.

\subsection{Semi-Structured Interviews}
Interview questions addressed similar topics as the survey, allowing us to gather more in-depth explanations. To summarize, we asked about: how participants generally used CaringBridge; how they used reactions across social media more broadly; their perception of the Heart reaction before and after the introduction of new reactions; and their opinions/thoughts/stories about the new reactions. (See supplemental materials for interview questions.) 

\subsubsection{Interview Recruitment}
361 out of 808 survey respondents opted in to be contacted for an interview. Of those who had opted in, we reached out to a small, random sample of participants a few weeks after the survey closed, balancing invites to authors and visitors. We continued recruiting until we had collected interviews from participants with a variety of spiritual beliefs and opinions about whether there should be fewer, the same, or more reactions on CaringBridge in the future. Participant IDs that begin with ``A'' indicate ``Author'' while ``V'' indicates ``Visitor.'' Table~\ref{tab:participants}b provides information about each interview participant.  

\subsection{Quantitative analytical approach}\label{sec:quant}
We used Python/Jupyter notebooks~\cite{jupyter_project_2021} to provide basic descriptive statistics on quantitative survey questions. At the time of the survey, 50.0\% of all respondents had seen the new reactions, and 29.3\% had received or used them at least once, meeting our goal to capture a large swath of opinions about the launch of new reactions on CaringBridge. All quantitative statements in the results represent the full set of 808 survey responses.

\subsection{Qualitative analytical approach}

We used Grounded Theory Method (GTM)~\cite{olson_curiosity_2014} to analyze the qualitative interview and free response data. Although Glaser and Strauss' original GTM technique suggests that researchers approach the data without bias or influence from prior theory~\cite{glaser_discovery_1967}, we adopted Charmaz' newer constructivist recommendations, which acknowledges researchers as subjective participants in the generation and analysis of the data, and which allows for the inclusion of ``sensitizing'' codes from prior literature, emergent inductive codes from the data, and iterated codes from later-stage analyses~\cite{mills_development_2006,charmaz_constructing_2014}. Sensitized to concepts from prior literature such as ``spiritual support,'' ``misinterpretation of emoji,'' and ``reactions as paralinguistic digital affordances,'' members of our research team first performed open coding on all free response survey questions and interview transcripts. Through a series of immersive team meetings, we then completed affinity mapping of the open codes using Miro~\cite{miro_miro_2021}, beginning with survey codes, and then interview codes, in order to determine axial themes across both data sources. We wrote memos throughout this process, iterating on codes and themes as they emerged. After determining our main themes, the first author then re-read all data, applied thematic codes, and selected quotes for presentation in the manuscript. 

Our methods of data collection afforded us the benefits of both a representative sample size (survey) and a deeper dive into the experiences of a smaller group of users (interviews). Although a quantitative content analysis (\textit{e.g.}, including an inter-rater reliability process and formal codebook) would be possible, we chose not to pursue this type of analysis because it was not necessary to address our research questions. Rather, in order to judiciously represent users' perspectives, we present summaries of qualitative insights from free response questions and interviews, alongside quantitative reporting on multiple choice or rating answers that demonstrate how these opinions generalize across our sample. [We also include notable examples of minority opinions held by approximately 10 or fewer respondents in brackets.] While we are unaware of prior work that uses this [bracketed] strategy, we offer this method of demarcation as a useful strategy for painting a complete and comprehensive picture of the data, without skewing readers' interpretations toward relatively rare exceptions.

\subsection{Limitations}

Our work must be interpreted in light of standard limitations applicable to surveys (\textit{e.g.}, selection bias, and the limitations of self-reported data) and interviews (smaller sample size, and the possibility of faulty recall of past events or inaccurate predictions of future feelings). These methods allowed us to first develop a strong qualitative understanding of users' perceptions; in our future work, we intend to address these limitations by triangulating our results from this paper with behavioral data collected from participants who opted-in to provide the email address associated with their CaringBridge accounts. In our discussion (sec.~\ref{sec:futurework}), we suggest research questions and directions for future work to guide these efforts, both on CaringBridge and other platforms with reactions. 

Additionally, our survey respondents are primarily female, white, and middle-aged. While this demographic distribution does not accurately represent the general public in the United States, it does accurately reflect the CaringBridge userbase~\cite{smith_i_2020,smith_what_2021}, as well as broader trends of healthcare workers, caregivers, and online health community users skewing female. Our sample is also majority Christian (67.2\%), however this closely aligns with a Pew Report showing that 70.6\% of the United States public is Christian~\cite{pew_religious_2014}. Future work should continue to evaluate reactions in other platforms and contexts, in order to understand more diverse types of users and populations.
\section{RQ1 Results: Differentiating CaringBridge reactions from other social media}

In order to provide a basis for understanding design intent, we begin by summarizing CaringBridge's rationale for adding new reactions. Next, we describe users' pre-existing perceptions of Heart, and how it is interpreted in its implementation as a standalone single-click reaction, especially given that \raisebox{-.5\mydepth}{\includegraphics[height=\myheight]{reactions/amp.jpg}} is a branded logo.

\subsection{Organizational rationale for adding new reactions}\label{sec:rationale}
Prior work suggests that CaringBridge users often struggle with finding the right words to express their support, and would appreciate technological assistance to \textit{``say the right thing.''}~\cite{smith_what_2021} In our preliminary conversation with CaringBridge employees, a C-Suite Executive said, \textit{``We see that when people look at that `write a comment' box, they become paralyzed, but they want to say, `hey I've been here.' ... When people don't know what to say, they want to push a button.''} This claim is supported by our data: Out of 371 visitors, 50.7\% often (165) or always (23) \textit{``don't know what to write in a comment,''} and 54.4\% often (137) or always (65) use a reaction when they don't know what to write. The Executive noted CaringBridge's overall mission to ensure no one goes through a health journey alone, and to provide users with as much support as possible. She shared four top-level motivations for expanding the reactions feature as: (1)~lowering the barrier for people to show their support with how they feel; (2)~expanding empathy for authors; (3)~meeting users' expectations for a modern social media platform; and (4)~strengthening the product and brand. 

\subsubsection{Why this particular set of reactions?}
A few years prior, CaringBridge had added an emoji picker tool to the commenting UI, allowing users to select from hundreds of emoji to mix within their comments. Since then, a Product Manager said that comments have included more and more emoji over time. In 2020, he observed a substantial and rising proportion of comments consisting \textit{entirely} of emoji, without any words--most especially sequences of prayer hands and other religious symbols. \textit{``That's one of the reasons we're like, ok, let's expand our emoji reactions, since people are just going around and doing that anyway.''} Given its importance to the brand, \raisebox{-.5\mydepth}{\includegraphics[height=\myheight]{reactions/amp.jpg}} would stay, while the addition of \raisebox{-.5\mydepth}{\includegraphics[height=\myheight]{reactions/pray.jpg}} was intended to facilitate a more convenient way to graphically express prayer. He explained that, since people are so accustomed to seeing smiley reactions across other social media platforms, two basic smileys would be included in the initial launch (\raisebox{-.5\mydepth}{\includegraphics[height=\myheight]{reactions/happy.jpg}} and \raisebox{-.5\mydepth}{\includegraphics[height=\myheight]{reactions/sad.jpg}}); more could potentially be added later. He also explained that the Product team initially \textit{``didn't want to box people in''} by forcing them to pick a single reaction. He recognized that people often have multiple emotions about a Journal update, however, \textit{``when pen got to paper, and we were figuring out how to do it in an intuitive way, it just became more complicated than not.''} Rather than extending the release date by months, they decided to stick with the limited-bar, since users already understood this reaction style, and to only pursue alternative reaction styles if users raised interest in that.%

Our evaluation of the product release reveals some ways in which CaringBridge's high-level objectives align with some users' needs and experiences. However, we also find that there are striking points of contention and disagreement. Throughout our results, our aim is not to argue for or against adding more reactions, but rather to synthesize aspects of participants' responses that demonstrate how the design decision impacts users' experiences. In the discussion, we then aim to unpack these complexities and trade-offs, providing guidance for future research and design.  

\subsection{User perceptions of Heart}
\subsubsection{Heart as a branding mechanism}
93\% of authors and 80\% of visitors indicated that they had previously seen, used, and/or received the Heart, indicating that respondents were highly aware of and accustomed to the Heart reaction. Furthermore, participants emphatically described a special feeling of connection and loyalty to CaringBridge as an organization, given the role it plays in helping them to streamline communication and coordinate meaningful support. 77.9\% of visitors and 80.5\% of authors agreed that Heart is \textit{``unique to CaringBridge, as opposed to other social media or emojis.''} Many users described how this specific organizational affiliation enriches their sense of supportiveness implicit in the use of the symbol as a reaction. For instance, S97 wrote, \textit{``I like this reaction [Heart] because I associate it with caringbridge. To me it is sending love in a special way compared to other social media emojis.''} [However, a few users expressed concerns that regardless of how much they love CaringBridge, it is inappropriate or disingenuous for \textit{any} branded logo to be used as a reaction, hence they refuse to use it.]

\subsubsection{CaringBridge reactions carry higher stakes}
Because CaringBridge is intentionally differentiated as a health support platform, in which people are facing life-threatening or even terminal illness, participants overwhelmingly highlighted how its reactions are of far higher stakes than on other social media platforms. Many participants noted some degree of similarity between Heart and the \textit{``default `like' button on Facebook''} (A4), however its functional ``default use'' operates differently on CaringBridge than elsewhere. For example, V5 described how \textit{``you can ‘like' a thousand things''} while mindlessly scrolling on Facebook or Twitter–but the same lazy reaction one might give to a silly cat video or a spilled cup of coffee is totally inappropriate for something as serious and difficult as a health crisis on CaringBridge. Aligning with prior literature on the Facebook Like~\cite{hayes_one_2016,scissors_whats_2016}, participants were also uncertain about what ``liking'' actually means. \textit{``Does that mean I agree? Does that mean I just saw it? You know, how do people interpret that?''} (A3)

\subsubsection{Heart as a ``general purpose'' reaction of acknowledgment and support}
\label{sec:perceptions_of_heart}
Many participants described Heart as a simple way to show that they had read a Journal update–yet on CaringBridge, this expression of acknowledgement is also consistently perceived as a support mechanism. For example, S512 wrote, \textit{``I like the beating heart visible at the bottom of the Journal update, for a quick acknowledgment that I read it,''} and many participants used a variety of words or phrases to describe what Heart means to them--\textit{e.g.}, love, care, prayers, a virtual hug, I'm here for you, I'm thinking of you. Furthermore, 79.0\% of visitors agree that Heart is a \textit{``way of saying I support you.''} Likewise, 85.6\% of authors ``feel personally supported'' when their Journal update receives a Heart, and 78.0\% would \textit{``rather have my visitors leave a Heart than do nothing.''} These results suggest that for most users, the CaringBridge Heart may be implicitly seen as more supportive than default reactions elsewhere, likely due to its special context, purpose, and branded look.

Participants often noted that as an expression of acknowledgement and care, Heart is a universally applicable reaction. No matter whether authors need to share good or bad news, Heart does not introduce emotional or cognitive dissonance. For example, V1 said, \textit{``whatever that person needs, they can take from just the heart''} while S236 wrote, \textit{``I think the heart reaction is the most useful and least likely to be misunderstood. You can't go wrong by using it.''} [At the same time, a few users shared that the uniqueness of the Heart design, and its inconsistency with the design of other emoji, made them unsure about what it meant, or they thought it meant that someone ``loved'' the content of the update, thus they won't use it on posts conveying difficult or upsetting information.]

\section{RQ2 Results: Transforming reactions into a complex decision-making interface}
\label{sec:rq2_results}

\resultssummary

There is no obvious consensus about whether adding more reactions (beyond the current set of four) is a good idea: 46.9\% want even more reactions added to CaringBridge, 45.2\% indicated that the new set of four is sufficient, and 7.9\% said there should be fewer (or no) reactions on CaringBridge. For example, one response from S325 details concerns about how increasing from a single Heart to a set of four reactions will alter their experience:

\begin{quote}
\textit{``I like that the heart is a simple, general reaction that says, `we're with you and thinking of you.' I think additional options just make things complicated for the well-wisher and the recipient. I do not want to try to decide which reaction is best for each situation as the well-wisher, or distinguish between what each person might mean with the emotion reactions as the recipient. Sometimes less is more and simple is sweet. I think this is one of those times.'' }
\end{quote}

Alluding to some of the concepts mentioned in this quote, we find that increasing the number of reactions substantially transforms the feature, and that adding even \textit{more} reactions could further accentuate this transformation. Figure~\ref{fig:resultssummary} summarizes four key transformations related to: interface complexity, functional ambiguity, semantic ambiguity, and issues of inclusion and exclusion. (See Appendix~\ref{sec:dataexamples} for example quotes of users' opinions or sentiments against or for adding more reactions, which also highlight these issues.)

\subsection{Interface complexity}
For visitors, increasing the number of reactions eliminates the simplicity of a general purpose Heart, and reacting must become a more active decision with a more complex UI/UX. Some participants felt that the more options there are, the more disorienting the decision-making experience may be. For example, V2 said, \textit{``If you have too many emojis out, like oh my God, is this what it is? Or is this what it is? So, like the number that Facebook has, I don't know that I'd want them to go much more because then I'd kind of be gone.''} Others felt unconcerned, such as S682, \textit{``Most apps have hundreds of emojis available. Why does Caringbridge have only four and why is this being treated like such a big deal? I'd say jump into the 21st century! People use emojis in their communication. That's how people write in informal speech and express themselves these days.''}

For authors, the interface for viewing who has reacted also becomes more burdensome and time-consuming, requiring them to click through four separate tabs and parse through four lists of names separately. Several participants expressed concerns that older users (a large proportion of the CaringBridge population) may struggle with these forms of complexity, both technically and socially, given that many of them may not have any other experience with reactions or use other types of other social media. For example, a number of participants were unfamiliar with ``hovering'' and could not figure out how to access the new reactions. 

\subsection{Functional ambiguity of smiley reactions (RQ2.a)}\label{sec:functionalambiguity}
New reactions introduce ambiguity around what reactions--particularly emotional face-based reactions--are functionally supposed to achieve. For example, S538 wrote, \textit{``I feel that the happy face and sad face can have multiple meanings behind why someone would select them. At times they can feel negative and off-putting versus supportive.''} Participants' comments indicated three distinct and sometimes clashing functions of reactions:

\begin{enumerate}
\item \textbf{Showing visitors' own emotional responses:} For example, V6 described how she decides which reaction to pick based on \textit{``what is closest to what it is that I'm feeling in response to what's been posted.''} While this may be unproblematic for good news, we received numerous concerns about Sad, because \textit{``you don't want to pile on the Sadness.''} (V5) Or, as V3 put it, \textit{``I probably would never use the crying face only because He knows we're all crying. The last thing he wants to do is make us sad.''} As an author, S551 also explained that, \textit{``It's difficult enough going through this without me worrying about how other people are dealing with it.'' }

\item \textbf{Empathy or solidarity:} Empathy was a commonly noted function, such that visitors should express the same emotion as what authors are feeling. For example, V6 said, \textit{``That tear [face] I think was a really appropriate thing to add. I think it's an expression of sympathy or empathy,''} while A4 said, \textit{``I tend to be more of a person that would interpret that in terms of solidarity.''}
 
\item  \textbf{Improve author's emotional state:} Some participants wrote that the purpose of reacting should \textit{always} be to encourage authors, which can imply diverging from the author's current emotional state in a reaction. For instance S246 wrote that, \textit{``When I'm feeling down it's nice to see a smiling face either in real life or an emoji,''} whereas others wrote that receiving Happy on a sad update would be frustrating and insensitive.
\end{enumerate}

\subsection{Semantic ambiguity of symbolic reactions (RQ2.b)}\label{sec:semanticambiguity}
Mirroring prior results on misaligned interpretations of emoji~\cite{miller_blissfully_2016,tigwell_oh_2016}, the inclusion of symbols as reactions also compounds upon semantic ambiguity in two important ways:

\begin{enumerate}
\item \textbf{The same symbol can refer to different items, gestures, or behaviors:} For example, some participants, such as S456, described the meaning of Prayer itself as \textit{``unclear - Is it thanks or prayers?''} [Other responses noted \textit{``namaste''} or \textit{``high five.''}] Even when the symbol is interpreted as prayer, different people can mean very different things by that. Some participants interpreted it as a flexible indicator of care or hoping for the best, whereas others have quite specific religious connotations in mind. V1 also pointed out that, \textit{``there's a difference between someone saying like, oh, I'm praying for you, we'll pray for you, but like actually doing it''}--highlighting a tension between single-click actions, and how they may or may not relate to behaviors they purport to represent.

\item \textbf{Overlapping meanings between different symbols:} Participants often used similar or equivalent phrases to describe Prayer and Heart with many participants explicitly including words like ``care'' and ``love'' to describe Prayer or ``prayer'' to describe Heart. Consequently, it may be unclear when to use one versus the other or what the true difference is between them. For example, A1 was unsure of why people vacillated between using the two: \textit{``I'm like, well why didn't you use the praying hands that other time? Why did you choose heart this time, not that time?''} Mirroring this same uncertainty from the visitors' perspective, V1 said, \textit{``I'm really wary of having to choose between, do I love this, or do I use praying hands? ... I prefer to just click it. I'm here. I saw it. I didn't have the option to choose anything else, and you didn't have the option to read into what I chose or didn't choose to click.''}

\end{enumerate}

\subsection{Inclusion and exclusion}\label{sec:inclusionexclusion}
Symbols like Prayer can both enhance feelings of inclusion within the represented group, while excluding others:

\begin{enumerate}
\item \textbf{Inclusion:} A large number of comments described the profound importance of \textit{``knowing that people are praying.''} (S196) For example, S90 described how, compared to Heart, Prayer connects to their spiritual beliefs in an elevated way: \textit{``To me the heart says, `yes, I care about you and your family' but the hands says, `I care about you, and I am asking for a higher level of help for you.'''}  Similarly, V2 connected this act of \textit{``calling upon a higher power to help with your healing''} with a feeling of deep and nourishing connection to her spiritual community.

\item \textbf{Exclusion:} A small set of non-customizable reactions cannot contain a representative and inclusive set of symbols. For example, S617 wrote that Heart is \textit{``neutral and inclusive,''} but highlighted two major concerns about racial representation (skin tones) and spiritual exclusivity. \textit{``The world is not white Christians. Unless you want to include emoji faces of every color, it excludes and does not represent. Unless you want to include emojis of every religion or non-religion, it excludes and does not represent.''} [We also received several spicy comments along the lines of \textit{``all or none''} (S420), with a handful of users explicitly claiming that they will abandon CaringBridge if they ever receive such an offensive reaction as Prayer.]

\end{enumerate}

In this section, we have demonstrated how the introduction of new reactions has transformed reacting on CaringBridge into a complex decision-making interface impacted by a variety of ambiguities and concerns. Next, we will describe RQ3 results to show how this transformation impacts support exchange more broadly.

\section{RQ3 Results: Shifting user expectations and assumptions}
Reactions comprise one type of interaction within a much broader paradigm of support exchange that includes observable on-platform behaviors (\textit{e.g.}, reactions, comments), responses mediated by other technologies (\textit{e.g.}, phone, email, social media), and offline responses. In this section, we first describe how the transformation of reacting affects authors' expectations and needs for supportive communication. Next, we compare authors' expectations to visitors' decision-making processes for how to respond.

\subsection{Authors' expectations and needs for supportive communication (RQ3.a)}

\subsubsection{Authors' need to reduce communication burdens.}
All author interviewees described the major need for starting a CaringBridge site as \textit{``streamlining the communications so that I could have one place to say things.''} (A7) When they began sites, they had very few expectations of what would come of it or how people would respond. As A1 put it, \textit{``I wasn't expecting a response. It was more like, I was hoping for consequences, like, people having access to that [information], you know, like don't call me and ask for the things that I just wrote.''} Yet as time progressed, most authors noted receiving an unanticipated outpouring of responses that became deeply meaningful to them, while simultaneously developing expectations and curiosities around which visitors would see their updates, and how they would respond.

\subsubsection{Authors develop the expectation that visitors should respond.}\label{sec:expectingresponse}
Despite an initial lack of expectations for responses, A1 said, \textit{``When I would see the notifications that somebody had reacted, I would look at the list of names, and it would be nice to know that they read this. ... And then it would kind of be like well, why, didn't everybody put hearts?''} This highlights a general expectation by authors that if people have visited an update, that they should respond in some way.\footnote{During site creation, CaringBridge prompts users to select a privacy level that enables (1) anyone, (2) only registered users, or (3) only invited individuals to view the Journal; (2) is most common. Authors who select (2) or (3) can access a feature that shows them a list of who has visited each update–a list that naturally lends itself to comparison with lists of which of these people reacted or left a comment.} For example, S347 wrote:

\begin{quote}
\textit{``The worst is when I pour my heart out and GOOD friends read it, and neither react nor comment. I'd love to receive a reaction from every person visiting. I think people don't understand how important it is for someone with an illness to feel cared for from readers. A simple heart goes a long way.  And - the absence of a heart really challenges me! It actually makes me sad - especially when I can see who has visited and not left a reaction or comment.  It makes me feel just bad.''}
\end{quote}

Given that most authors expect visitors to respond in some way, the following sections describe how an increase in the number of available reactions shapes new expectations for how visitors' responses should look, first focusing on reacting, and next examining the dynamics of reacting versus commenting.

\subsubsection{Author expectations of multiple-option reacting}\label{sec:authorexpectations}
Increasing the number of reactions impacts authors' expectations around reacting in two ways:

\begin{enumerate}

\item \textbf{Appropriate alignment to content.} Authors expect visitors to align their selection of reaction with the content of their updates. For example, S671 wrote, \textit{``The entry I made today was a bit of levity amongst all the deadly seriousness of my cancer journey. Honestly, I would be offended or concerned if someone responded with the sad face today. If today's post had been a sad one, then the laughing face would have rankled and been unsuitable.''}  Users also pointed out how risks of confusing or frustrating misalignment are higher when there are mixed emotions or events in a given Journal update, such that it's unclear which part of the update the reaction applies to. 

\item \textbf{Which specific reactions will be used.} Authors emphatically described why they would or would not want to receive specific reactions. For example, many Christians wanted and expected to receive Prayer, whereas for S675, an atheist, Prayer \textit{``feels like a tiny jab every time.''} Another example demonstrates the disappointment that S685 feels, because she expects that she may never receive Happy: \textit{``I would love to get this [Happy] reaction because it would mean that people see something positive in his recovery. I'm afraid I will never see this reaction since I now believe he will likely die within the next year.''}

\end{enumerate}

\subsubsection{Author expectations of reactions versus comments}\label{sec:reactionsvscomments}
There exists a notable split in authors' opinions of reacting vs. commenting. For instance, 47.6\% (n=208) of authors said that receiving comments is more important to them than receiving reactions. In contrast, 49\% (n=215) said they were equally important, while only 14 indicated that reactions are more important. We identified three main factors influencing these opinions:

\begin{enumerate}
\item \textbf{Meaningfulness.} Many authors expressed that comments are more meaningful to them than reactions because of the time and thoughtful consideration that goes into typing out words rather than clicking a button. Some users even went so far as to express concerns about the growing illiteracy of society, wishing people could write at least a few simple words, rather than \textit{``poking a cartoon face.''} (S621) However, other authors acknowledged how busy their visitors are in their daily lives, granting grace to reactions as a visually expressive \textit{``form of short hand''} (S288) that communicates care in a different way. (See full quotes in Appendix~\ref{sec:dataexamples}.) 

\item \textbf{Efficiency.} CaringBridge sites are often maintained by overburdened caregivers running on limited time, sleep, finances, etc. It is often the case that their beloved family member is incapacitated, thus they need to present a condensed version of visitors' responses at the bedside. Rather than reading dozens of repetitive variations on phrases like \textit{``we're praying for you,''} these users commented that they prefer Prayer reactions to save on time and cognitive strain. Some visitors also recognize the draining potential of repetitiveness. For example, S596 likes Prayer for \textit{``when I want to express that I care, but don't want to keep repeating myself with `prayers and thoughts.'''}

\item \textbf{Stage in health journey.} Some authors observed social and temporal dynamics in how visitors leave reactions vs. comments. For instance, one author described how the volume of both reactions and comments seemed to swell during moments of extreme intensity or uncertainty in the health journey, as people who would not typically respond became more active. Responses tapered off for less urgent or more routine updates, skewing toward a higher proportion of reactions (or non-response). Moreover, when a health issue is chronic or persists over years, A5 noted, \textit{``there's got to be a friend-and-family fatigue,''} making it difficult to keep commenting on every update. Building on this idea, A2 provided a subjective estimate that 80\% of visitors react and about 20\% comment on any given update, but \textit{``it's not necessarily the same 20\% continuously.'' }

\end{enumerate}

These considerations highlight important tensions between the meaningfulness, efficiency, and timing of responses. Because of increased complexity of the reaction options and interface, some authors noted concerns that adding more reactions could reduce comments, or cause a higher proportion of non-responses from visitors who become too confused by the interface itself or the cognitive evaluations needed to decide on a response type. Thus, we next describe visitors' perceptions of how new reactions impact their decision-making processes.

\subsection{Multi-option reacting impacts visitors' decision-making processes (RQ3.b)} We identified four ways in which increasing the number of reactions can impact the way visitors choose to respond to Journal updates.

\subsubsection{Amount of cognitive effort required.}\label{sec:cognitiveeffort} At the beginning of section~\ref{sec:rq2_results}, a quote from S325 demonstrates concerns about how increasing the number of reactions increases cognitive burden, both in terms of selecting a reaction and in reasoning about how authors will interpret that selection. [A few users also expressed concerns about the possibility of accidental selection of the wrong reaction.] On the other hand, some authors pointed out that the effort of having to choose between reactions \textit{``would feel a little bit more personal''} (A4) or shows more thoughtfulness because \textit{``it does feel like they like stopped and had to think,''} (A1) unlike a single-click response.  Some visitors, most especially those who had previously been authors, feel that cognitive effort of writing a comment, as opposed to reacting, is required for a response to be meaningful, such as S621 who considers it a \textit{``sacred obligation''} to comment \textit{``if someone is sick enough to have one of these accounts set up.''} 

\subsubsection{Relational closeness.}\label{sec:socialcloseness} Participants described how relational closeness impacts their decision to react or comment. For example, S727 said, \textit{``I am more likely to use the emoji's with someone I do not know as well.''} A7 described how social closeness could even impact which specific reactions visitors might feel comfortable using, \textit{``I think for a lot of people who don't regularly use social media that the heart emoji might feel too personal. The same person might not sign something like `love' when writing to me, especially a coworker or something like that.''}

\subsubsection{Social dynamics of publicly visible responses.}\label{sec:socialdynamics} Visitors described how the knowledge that others will see their responses can impact their behavior. For example, S189 struggled to leave comments for this reason, opting for a reaction instead. \textit{``Sometimes I don't know how to write something to my friend \& especially something so personal that so many others will read publicly. The heart reaction is helpful as I don't want to ALWAYS be commenting \& monopolizing things. It is a fine line.''} On the other hand, comments and reactions that are already visible to the community can also impact how people respond. As V6 pointed out, \textit{``You see how the community, how everybody else in the situation is responding and then that colors your response.''} Similarly, A7 wondered about visitors who viewed her posts later, once many comments had already been left by others. She imagined that these latecomers might react rather than comment because they \textit{``could say the same thing, but like what else am I gonna say?''}

\subsubsection{Visitors' personal characteristics or personality traits.}\label{sec:personalcharacteristics} Some participants described how their personal characteristics greatly inform their choice of response. For example, some participants attributed their appreciation for or dislike of emoji to their age, and how emoji have or have not been socialized among members of their generation. For example, S906 (who selected the ``55-64 years old'' age bucket) wrote, \textit{``Having too many `reactions' can lead to miscommunication...especially for us older folks.''} Other participants focused on intrinsic personality traits. For example, V1 said, \textit{``I think I'm just not a comment person. Some people are just comment leavers, you know, like good for you. And I'm just like, okay, I'm here for you–click!''} Similarly, V6 said, \textit{``I'm a very introverted person, so I'm more interested in taking in information.''} Instead of responding on CaringBridge in any way, V6 said they call the patient directly to leave a voicemail, \textit{``because it's a personal communication. It's not for public display.''}

\phantom{force a blank line}

As the last comment from V6 alludes to, our interviewees shared many stories about how the proliferation of information via their CaringBridge site also caused touching offline responses, such as texts, calls, cards, flowers, and instrumental support. Some participants shared stories of even larger community-organized efforts after making a difficult post on CaringBridge. For example, A5 shared the story of an adult patient (who loved aloha shirts) receiving a wave of pictures from friends wearing aloha shirts in his honor, and A7 told a story about a child with cancer receiving a set of tiger masks from his baseball team (The Tigers), to be worn by the team the next time the child could attend a game. 

As A3 put it, all of these different manners of responding--reacting, commenting, or responding offline--can be \textit{``encouraging in different ways.''} Thus, in our discussion, we next unpack how our results can inform the design of supportive online communities and systems, in order to best align the types of support that visitors can provide with the needs and expectations of patients and caregivers.

\section{Discussion}
In January 2021, CaringBridge transitioned its single-option, single-click Heart \raisebox{-.5\mydepth}{\includegraphics[height=\myheight]{reactions/amp.jpg}} reaction into a limited-bar reaction-style, now enabling users to select from Heart~\raisebox{-.5\mydepth}{\includegraphics[height=\myheight]{reactions/amp.jpg}}, Prayer~\raisebox{-.5\mydepth}{\includegraphics[height=\myheight]{reactions/pray.jpg}}, Happy~\raisebox{-.5\mydepth}{\includegraphics[height=\myheight]{reactions/happy.jpg}}, or Sad~\raisebox{-.5\mydepth}{\includegraphics[height=\myheight]{reactions/sad.jpg}}. CaringBridge's organizational aims for this product launch included lowering the barrier for people to show their support with how they feel, expanding empathy for authors, meeting users' expectations for a modern social media platform, and strengthening the product and brand. This paper has evaluated users' perceptions of the launch, enabling us to now discuss, relative to CaringBridge's goals---\textit{how did it go?}

\subsection{Designing reactions for health-specific platforms \textit{versus} mainstream Online Social Networks (OSNs)}
On Facebook, single-option, single-click Liking means many different things to many different people~\cite{hayes_one_2016,scissors_whats_2016}, and people often find it particularly mindless or reactive relative to other platforms~\cite{hayes_one_2016}. CaringBridge offers a striking point of comparison. As a single-option, single-click PDA, most users experience Heart~\raisebox{-.5\mydepth}{\includegraphics[height=\myheight]{reactions/amp.jpg}} simply as general-purpose reaction of support and acknowledgement. As suggested by prior design ideations for a socially supportive ``Hug''~\cite{ahmadi_designing_2016} or ``Empathy''~\cite{choi_you_2022} reaction, Heart can be universally applied to any Journal update, without much risk of misinterpretation or offense. Given the life-and-death nature of CaringBridge's context, and users' sense of loyalty, appreciation, and warmth toward the brand, both authors and visitors associate Heart with more meaningful support than Likes, with some users sharing that they chose CaringBridge specifically to get away from the vapid nature of other OSNs. Thus, our work reveals a significant tension between CaringBridge's goal to meet users' expectations of modern OSNs and the reality that PDAs on other OSNs can be overly complex or offputting in the context of a health crisis. Should CaringBridge \textit{keep it simple}, or should CaringBridge \textit{keep up with the Joneses}, so to speak?

\subsubsection{Keeping it simple: Selecting one single-option, single-click reaction}
One possible interpretation of our results suggests that CaringBridge should maintain a single-option, single-click reaction-style as the simplest possible affordance. Most authors are middle-aged or older adults. Many have moved into a later period of life, in which they and their friends need to regularly confront the onset of diseases like cancer, cardiovascular failure, or neurological degeneration. At that stage, users may not care as much about modern, conveniently expressive OSN affordances as much as they prefer ease-of-use and in-depth responses (\textit{e.g.}, comments or offline responses). In this case, is Heart a good single-option? 

Our results suggest that Heart~\raisebox{-.5\mydepth}{\includegraphics[height=\myheight]{reactions/amp.jpg}} is an excellent single-option single-click PDA option for most CaringBridge users. It capitalizes upon strong community perceptions of CaringBridge as a safe space of support and love, differentiating itself in both design and meaning from other emojis and platforms. It is not explicitly a spiritual symbol but nonetheless carries spiritual connotations for some users (\textit{e.g.}, overlapping semantic interpretation of ~\raisebox{-.5\mydepth}{\includegraphics[height=\myheight]{reactions/amp.jpg}} as loving prayer); this flexibility makes it welcoming to anyone. As opposed to a reaction-free interface, it provides the benefit of a supportive acknowledgement for people who don't know what to say without overly complicating the interface. In designing single-option, single-click PDAs, other health-support platforms should ensure that their choice of symbol: (1) appeals to that community's particular zeitgeist or brand; (2) clearly evokes supportive intent; and (3) can be universally applied without risk of misinterpretation or offense. In general, smileys cannot meet all of these criteria, whereas heart- or other symbol-based designs may be a good starting point.

We asked users whether the new set of four is sufficient, or if there should be fewer or more reaction options. Realistically, only 7.9\% of respondents indicated that there should be fewer (or no) reactions, thus it would contradict the majority to return to the single-option, single-click reaction-style. As the childhood adage goes, perhaps \textit{``no takebacks!''} applies here. However, our work does suggest that health-focused platforms should seriously consider whether or not expanding their reaction-style is truly the best design direction, even if \textit{``everybody else is doing it.''} Sometimes, lower tech solutions are better for users, and the implication is \textit{not} to design something fancier~\cite{baumer_when_2011}. Assuming that we \textit{will} serve users a limited-bar, however, how should designers select the right \textit{set} of options for a health-focused platform? 

\subsubsection{Keeping up with the Joneses: Selecting the right \textit{set} of reactions in a limited-bar}
In our study, 45.2\% of users indicated that CaringBridge's new set of four reactions is sufficient. These responses tended to express appreciation for the expressive new options, or a bit of ambivalence---the set seems fine at a glance, and comprehensive-enough-ish, whereas adding more reactions would become overwhelming and burdensome. A comparable, though slightly higher, proportion of 46.9\% want more reactions. Reflecting a similar conclusion from Wisniewski's study of Facebook's reactions launch~\cite{wisniewski_happiness_2020}, these participants sometimes mentioned the \textit{incompleteness} of the set for expressing all emotions. (Some highlighted a need for ``Anger'' in particular---\textit{e.g.}, for when chemo doesn't work---whereas others said ``Anger'' would be wholly awful and inappropriate on CaringBridge.) 
Rather than focusing on emotions, however, other users highlighted the need for reactions that are conceptually aligned with health journeys. Drawing upon nature- or gesture-based symbols, these users suggested options like flowers, rainbows, sunshine, celebration or clapping hands, hugs, care (directly from Facebook), candle, thinking of you, or even medical symbols, \textit{e.g.}, cancer ribbons, chemo, remission.

Given how elusive and confusing emoji can be in their own right~\cite{miller_blissfully_2016,miller_understanding_2017,miller_hillberg_what_2018,tigwell_oh_2016}, our work highlights how using emoji as reactions can compound their potential for misinterpretation or mismatched understandings. CaringBridge's reactions launch not only increased the complexity of the interface, but also introduced new \textit{functional ambiguities} regarding what emotion-based smiley reactions are supposed to achieve, new \textit{semantic ambiguities} around symbols and what they refer to or when to use which, and most importantly new concerns related to \textit{inclusion and exclusion}. Consequently, designers of limited-bar reaction-styles need to direct attention toward \textit{reducing ambiguity} and \textit{improving inclusivity} of the reaction symbols available. We next suggest design implications for achieving these aims. %

\textbf{Reducing ambiguity:} Whereas CaringBridge aimed to allow visitors to express how they feel and expand empathy for authors, we have shown that these aims are sometimes in conflict with each other or with users' interpretations of the feature. Thus, we ask readers to consider---what \textit{form} of expression is most relevant and needed within the specific paralinguistic affordances of reacting? Do people really need to express \textit{emotions} through PDAs on CaringBridge? Or in this specialized context, is emotional expression perhaps more suitable within comments, where users can combine text with any smileys or other emoji that they so choose? We posit that Facebook's initial selection of \textit{emotion-based} reacting is only one possible ``palette'' of expression---a particular palette that Facebook has since taken advantage of to hook users and drive profit.\footnote{For example, Facebook weights Anger~\raisebox{-.5\mydepth}{\includegraphics[height=\myheight]{reactions/fb_angry.png}} reactions five times more heavily than Likes~\raisebox{-.5\mydepth}{\includegraphics[height=\myheight]{reactions/fb_like.png}}, driving these posts higher in people's feeds to generate more engagement with the platform~\cite{merrill_five_2021}.} This initial selection may have overly biased other platforms, such as CaringBridge. As a first step, we suggest that a palette of health-promoting \textit{gesture-based} expression may be better for CaringBridge. For example, replacing smiley reactions with gesture-based reactions may help to reduce ambiguity in at least three ways: (1) It removes the functional ambiguity of emotion-based reacting, along with the possible errors and misalignments that could occur, such as ``heaping on the sadness'' with a sad face symbol when the patient would never want that, or trying to encourage an author with a smile, when that smile might feel intensely insensitive. (2) It removes issues with understanding the gradation of emotion associated with a given smiley~\cite{miller_blissfully_2016}, or a given emotion only applying to a specific part of a post. (3) It represents possible supportive actions visitors \textit{would} do, were they co-located (\textit{e.g.}, hug) or even \textit{can} do, despite being far away (\textit{e.g.}, pray, light a candle). For these reasons, we suggest that communicating supportive gestures may be more appropriate than communicating emotion.

\textbf{Improving inclusivity:} We observed two dimensions of inclusivity related to race and spiritual beliefs. First, if smileys and/or hands are included as reactions, this raises crucial issues about racial inclusivity. Prior work on emoji shows that people who opt to alter skin tones from the default yellow almost always choose a skin tone similar to their own~\cite{robertson_emoji_2020}, suggesting that self-representation and identity are crucial in PDA usage. Major systematic, race-based inequities already exist in healthcare and its affiliated suite of informatics technologies~\cite{veinot_good_2018}. It is inappropriate to increase barriers to usage or adoption by users who do not affiliate with the yellow skin tone. LinkedIn's colorful set of reactions skirt racial distinction, with skin tones of blue, green, purple, and aqua--the only yellow is a light bulb. Thus, one solution appears to be: select colors that do not resemble \textit{anyone's} actual skin tone.\footnote{Tattoos excepted, and appreciated. ;-)} Another solution would be, like emoji, to enable users to select the skin tone of their reactions. Second, most users love and expect to use the Prayer~\raisebox{-.5\mydepth}{\includegraphics[height=\myheight]{reactions/pray.jpg}} reaction, with many respondents citing the profound importance of knowing that their loved ones are praying, and being prayed for---strengthening the growing body of evidence in CSCW demonstrating the importance of digital prayer and spirituality~\cite{smith_i_2020,smith_what_2021,kaur_sway_2021,oleary_community_2022}. However, different religious and spiritual traditions pray using different gestures and body positions. Whereas some participants expressed fluidity in their interpretation of the symbol, with acceptance that it generally indicates care and positive intent, others were enraged to see it appear on \textit{their} Journal. A space that used to feel safe now felt like a space where people could digitally prod them with reminders of religious imposition.

\subsubsection{Considerations for spiritual inclusivity}\label{sec:spiritualinclusivity}
Because spirituality is of such core importance to the vast majority of users, we strongly endorse the value of designing technospiritual affordances for prayer. Yet the matter of spiritual inclusivity is tricky to solve--particularly given the constraints of a limited-bar reaction-style. Including both religious and non-religious language, \citeauthor{smith_i_2020} define prayer support broadly to include \textit{``prayers, spiritual blessings, positive karma, good juju, warm thoughts, etc.''}~\cite{smith_i_2020} In the context of reactions, an ideal solution would be \textit{one} symbol that both generically references \textit{any} of these forms of prayer support and simultaneously evokes a concrete connection to each user's own specific spiritual belief system~\cite{smith_what_2021}. But this may be difficult or impossible to achieve in practice. For example, rather than praying hands~\raisebox{-.5\mydepth}{\includegraphics[height=\myheight]{reactions/pray.jpg}}, a new design consisting of the literal word ``pray'' within an icon (\textit{e.g.},~\raisebox{-.5\mydepth}{\includegraphics[height=\myheight]{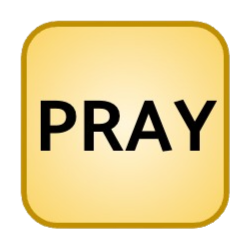}}~) could eliminate tensions related to religious differences in gestures associated with prayer. Yet this choice would reduce the \textit{non-verbal} expressive benefits of PDAs, and it also doesn't help users who are specifically offended by the concept of prayer. Likewise, imagine that a reaction like ``Thinking of You''~\raisebox{-.5\mydepth}{\includegraphics[height=\myheight]{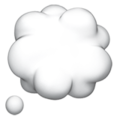}} were introduced---this is unlikely to offend anyone, but it is also unlikely to evoke a specific connection to the sacred. We offer three solutions:

\reactionpicker

\begin{enumerate}
    \item \textbf{Opting-out:} One obvious way to fix this problem is to retain Prayer~\raisebox{-.5\mydepth}{\includegraphics[height=\myheight]{reactions/pray.jpg}}, since the majority love it and it is evocative of the extremely commonly used ''folded hands'' emoji~\cite{daniel_most_2021}, but enable authors to opt-out of this particular reaction (or any others they don't want)---several participants readily suggested this solution. Moreover, prior work in the ethics of digital well-being suggests that opting-out is an important way to preserve human autonomy and promote the adoption of technology~\cite{calvo_supporting_2020}.
    \item \textbf{Choose-your-own spiritual symbol:} Another way would be to specially reserve one reaction in the limited-bar as a technospiritually-oriented affordance, with~\raisebox{-.5\mydepth}{\includegraphics[height=\myheight]{reactions/pray.jpg}} as the default---however, for this one reserved spot, enable authors to select from a far more comprehensive and representative set of religious and spiritual symbols, including symbols desired by agnostics or atheists.
    \item \textbf{Reaction-picker:} Allowing authors to choose their \textit{own} limited-bar from a pre-selected set would be more complex but more personalized to authors' desires. For example, CaringBridge could retain Heart~\raisebox{-.5\mydepth}{\includegraphics[height=\myheight]{reactions/amp.jpg}} as its default PDA beneath Journal updates to maintain consistency of the UI appearance across all sites, but allow authors to select up to $n$ other custom-branded reaction graphics to appear in the hover menu. See Figure~\ref{fig:reactionpicker} for a suggestive mock-up interface. We offer this mechanism as a compromise between the benefits of constraining options in a limited bar (\textit{i.e.}, limiting the cognitive decision-making burdens on visitors) and the benefits of the free-for-all (\textit{i.e.}, providing authors with some options without overwhelming them with \textit{too many} options).
\end{enumerate}

We add a note of caution that, even though customization of limited-bars could address issues for tech-savvy users, very few people alter default settings. Moreover, the more ``customizable'' a limited-bar becomes, the lesser its merits seem, esp. when compared alongside the free-for-all reaction-style (such as on Slack, where users can input \textit{any} emoji as a reaction). The free-for-all helps with inclusivity, albeit at the expense of increasing cognitive load, complexity, and ambiguity, in ways that might be too frustrating for CaringBridge's current userbase. Consequently, another solution that improves the spiritual inclusivity of reacting, while preserving the constrained benefits of the limited-bar, would be to remove prayer from reacting altogether, and instead design separate, new, technospiritual affordances for prayer, such as interactive real-time visualizations~\cite{kaur_sway_2021} or prayer walls~\cite{oleary_community_2022}. (Similar to reactions, any alternate prayer affordances would also benefit from an opt-out mechanism.) How and where to best communicate prayer within user interfaces remains a fascinating and complex open question for future work.

\subsubsection{Concluding design recommendations for CaringBridge reactions}
Pragmatically, at CaringBridge, the best solution \textit{today} is likely to be the simplest one for both users and developers.\footnote{Optimal \textit{future} solutions may continue to favor simplicity. Or, as today's younger generation ages, they may retain engrained expectations for health-specific platforms to adopt more complex CMC affordances and practices, such as the free-for-all reaction-style, or new forms of reacting altogether.} That could mean either returning to the single-option, single-click reaction-style, or refining the set of options within the current limited-bar. For example, we suggest that a better set of reactions for CaringBridge might include Heart~\raisebox{-.5\mydepth}{\includegraphics[height=\myheight]{reactions/amp.jpg}} and Prayer~\raisebox{-.5\mydepth}{\includegraphics[height=\myheight]{reactions/pray.jpg}}, alongside custom-branded graphics similar to 
Hug~\raisebox{-.5\mydepth}{\includegraphics[height=\myheight]{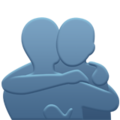}}~, Flowers~\raisebox{-.5\mydepth}{\includegraphics[height=\myheight]{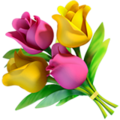}}~, Candle~\raisebox{-.5\mydepth}{\includegraphics[height=\myheight]{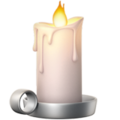}}, and Clap~\raisebox{-.5\mydepth}{\includegraphics[height=\myheight]{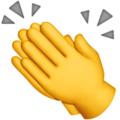}}---ideally with non-yellow skin tones \textit{and} the ability to opt-out of any of these. This collection of gesture-based reactions would enable people to express supportive gestures across both difficult and celebratory moments. Moreover, it does not preclude users from using emotion-based emoji in their comments, where they can possibly clarify their meaning with accompanying text~\cite{alshenqeeti_are_2016,zhou_goodbye_2017,daniel_emojis_2018,miller_understanding_2017}. We note that there remains semantic ambiguity (\textit{e.g.}, sec.~\ref{sec:semanticambiguity}) between some of these symbols. Empirical investigations, with methods such as those used by~\cite{miller_blissfully_2016,tigwell_oh_2016}, could be used to evaluate an even larger set of prospective reactions, identify how users perceive the meanings of each symbol, and ultimately select a subset which results in the least overlap.

Beyond reaction-styles, we next discuss how the needs and expectations of CaringBridge authors and visitors can help to inform future CMC affordances for health more broadly.

\subsection{CMC affordances for addressing the specialized needs of Online Health Community (OHC) users}
Authors need explicit communicative actions to feel supported, and they need them consistently across health journeys which may endure for months, years, or decades. As suggested by the Optimal Matching Model of Stress and Social Support~\cite{mattson_health_2011}, the type of supportive communication provided by visitors should match authors' needs. Therefore, we next provide suggestions for how OHC designers should focus efforts on facilitating supportive CMC that is specifically needed and expected by authors. Given how difficult it can be for visitors to know what to say or what type of help to provide~\cite{smith_i_2020,smith_what_2021}, we also consider opportunities to improve the clarity, ease of use, and utility of affordances for support provision by visitors.

\subsubsection{Empowering authors to shape expectations of feedback from visitors}
People's expectations for feedback from their audiences shape their usage of OSNs like Facebook and Instagram~\cite{dumas_lying_2017,grinberg_understanding_2017,hayes_its_2016}. Our results demonstrate that this holds true for CaringBridge as well, albeit in unique ways. Although authors told us they did not initially have expectations, they developed the idea that visitors \textit{should} respond somehow---\textit{i.e.}, just viewing the Journal update is insufficient. In fact, when authors see that someone has visited \textit{without} leaving a reaction or comment---a special feature available on CaringBridge \textit{v.s.} mainstream OSNs---this can result in negative and damaging feelings, esp. in close relationships. A brute-force solution would be to automatically apply Hearts whenever people visit updates, however perceived automaticity decreases perceived supportiveness~\cite{carr_as_2016}. Given that authors need \textit{genuine} care, removing intentionality from reacting would undermine its purpose. Rather, we suggest that OHC designers build mechanisms to improve authors' ability to receive the types of responses they want, and visitors' awareness of the importance of responding. 

\begin{figure}[H]
    \centering
    \includegraphics[width=0.6\textwidth]{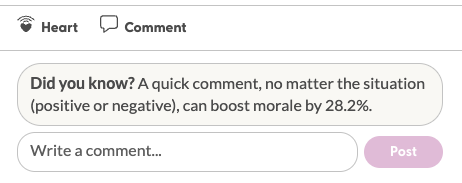}
    \caption{Screenshot captured in July 2022 of messaging displayed to prompt CaringBridge visitors to comment.}
    \label{fig:boostmorale}
\end{figure}

As mentioned above in sec.~\ref{sec:spiritualinclusivity}, prospective control mechanisms include allowing authors to customize sets of reactions, opt-out of certain reactions, or even to turn off the entire feature. Likewise, the ability to turn off commenting could be situationally or temporarily useful---\textit{i.e.}, to remove overwhelming amounts of cognitive processing when caregivers are too overburdened, want relief from tricky social dynamics that can play out in comments, or simply don't want input on a particular post. Additional strategies could be more \textit{instructive} to visitors. For example, Figure~\ref{fig:boostmorale} shows a static prompt that is currently displayed to CaringBridge visitors below Journal updates, encouraging them to leave comments.

This messaging would be improved if it included tactical details related to specific authors' own needs and preferences for supportive CMC. We suggest the concept of \textbf{author-configured nudging}, as a form of social nudging~\cite{bergram_digital_2022} in which authors could indicate a set of CMC preferences (\textit{e.g.}, answering a quiz of 1-5 preference eliciting questions), resulting in more customized text presented to visitors. Imagined outputs include copy (and a prospective bouquet-of-flowers reaction option, ~\raisebox{-.5\mydepth}{\includegraphics[height=\myheight]{emoji/bouquet_1f490.png}}), such as:

\begin{itemize}
\item \textbf{It would mean a lot to [Name], to know that you are praying!} Send her some ~\raisebox{-.5\mydepth}{\includegraphics[height=\myheight]{reactions/pray.jpg}}, or leave a comment.
\item \textbf{[Name] would really appreciate your comments below!} Words say it best, even if you only have a few. %
\item \textbf{Now is a time for quiet.} Please send ~\raisebox{-.5\mydepth}{\includegraphics[height=\myheight]{emoji/bouquet_1f490.png}} and~\raisebox{-.5\mydepth}{\includegraphics[height=\myheight]{reactions/amp.jpg}}. 
\item \textbf{[Name] could use a pick-me-up.} Share a hopeful message or a pleasant memory you have with her. 
\end{itemize}

As implied by the latter two examples, we further suggest that this concept of author preference elicitation be continually applied over time, to account for an author's needs changing over the extended course of their health journey. Likewise, our author participants have observed visitor behavior shaped by the stage of the health journey: swelling during intense moments in the journey and tapering off during less intense moments. Such health journey-induced ebbs and flows of support may or may not align with author needs. For example, an author might crave more support than perceived during a prolonged period of ``non-urgency.'' Thus, author-configured nudging may also be a unique opportunity to shift control to authors instead of their health journey commandeering. %

In addition to fostering alignment between author needs and visitor behavior, this suggested affordance carries potential for reducing known challenges of support provision~\cite{smith_i_2020,smith_what_2021} %
by removing guesswork for visitors. Any cognitive effort that may have been spent deciding how to respond can thus be redirected to the response itself. Furthermore, given our finding that personal characteristics or personality traits can influence visitor behavior (\textit{e.g.}, ``not a comment person''), greater awareness of author preferences created by such a feature could nudge visitors to ``make an exception'' to their preferences, and comment anyway (or vice-versa--\textit{i.e.}, react when their preference is to comment).  Finally, explicit awareness of ongoing author needs could help counter visitor ``fatigue'' experienced over an extended health journey: the nudges not only encourage but also \textit{validate} prolonged support provision efforts.

\subsubsection{Enabling private responses by visitors}
OHC designers should also consider enhancing the commenting feature by enabling visitors to make their comments either: public to anyone else visiting, or private and viewable only by authors---another idea suggested by several participants. The basic premise of an OSN is to afford communication amongst members by providing interaction mechanisms (\textit{e.g.}, PDAs and comment boxes), and then displaying the interactions that have taken place. In mainstream OSNs, this functionality is fundamental, however users can also send direct or private messages (a function that CaringBridge does not currently offer). In fact, CaringBridge \textit{needs} to maintain its one-to-many mass communication style, in order to meet authors' major need for streamlining communication. Adding direct messaging would likely add many new threads of private conversation for authors to keep track of, and this seems antithetical to the purpose of CaringBridge, whereas it would also add another layer to visitors' decision-making. (\textit{E.g.}, Do I react and/or comment publicly? Do I \textit{also}, or \textit{instead}, send a private message directly to the author?)

We know from prior work that people's decisions to respond (or not) are influenced by personal and social considerations, as well as platform-specific affordances~\cite{andalibi_responding_2018}, and that more intimate communication and support is shared privately rather than publicly~\cite{yang_channel_2019}. In the context of a health platform, should \textit{all} support exchange be on display? For some of our participants, the social context imposed additional pressure on the experience of responding: knowing that the response will be on display, and/or a lack of \textit{original} words can discourage commenting. A straightforward solution would be to enable privacy of individual or \textit{all} comments. Both of these solutions could afford more genuine responses, while also maintaining the explicit connection between comments and a given Journal update, without spinning off separate, cognitively-taxing conversations. Essentially, we are suggesting a shift to more authentic one-on-one, visitor-author support exchange (while maintaining the crucial benefit of mass communication of health updates by the author in the opposite communicative direction).

\subsection{Future Research}\label{sec:futurework}
Even though users mostly \textit{accept} the new reactions feature, our work nonetheless questions whether expanding into a limited-bar was actually a good idea for CaringBridge given the qualitative and conceptual tensions we have highlighted. One critical high-level question that can \textit{quantitatively} refine future recommendations is: Did the feature release contribute to increasing or decreasing the total number of supportive interactions exchanged? If it \textit{decreased} meaningful commenting behavior and/or did \textit{not} increase the total amount of support received by authors, this information would strengthen the case for nixing the launch and returning to a single-option, single-click PDA. If it lowered barriers to people expressing support, thereby increasing the total number of supportive interactions received, then perhaps the new complexities and issues introduced by the feature are tolerable, in service of the platform's ultimate goal to maximize support for users.

As the next phase of this work, we intend to use interaction data collected in-the-wild on CaringBridge to study how the reactions release impacted actual user behaviors compared to their stated preferences and expectations. Which reactions do users tend to use, and when do they use them? How does this impact other CMC affordances, such as commenting? In support of these future efforts, we append a list of quantitative research questions inspired by our qualitative results (see Appendix~\ref{ref:futureRQs}). We encourage researchers to study these types of questions in other contexts as well. It will be interesting and informative to develop contextual understandings of different reaction-styles across many different OSNs and platforms. For example, we hope future researchers will conduct studies of the free-for-all reaction-styles on platforms like Slack and Discord in order to develop a better understanding of how emoji, reactions, and other visual forms of CMC are now shaping influential aspects of workforce health~\cite{kaur_i_2022}, communication norms, and possibly other relational, health, and quality of life outcomes. Moreover, rather than prioritizing brisk, convenient forms of CMC, recent work at CSCW highlights the importance of \textit{effortful} communication to facilitate authentic and meaningful user interactions~\cite{zhang_auggie_2022}. More specifically, \citeauthor{zhang_auggie_2022} provide valuable design recommendations for: (1) designing \textit{approachable} effort through masking the heaviness of effort, reducing the skills needed for content creation, and facilitating asynchronous CMC; and (2) creating \textit{meaningfulness} through techniques such as personalization, bridging physical and digital contexts, and explicitly revealing the effort that went into crafting CMC expressions~\cite{zhang_auggie_2022}. With the rapid onset and public adoption of generative artificial intelligence in 2023, our work joins~\citeauthor{zhang_auggie_2022} in emphasizing the importance and urgency of investing true effort, care, and authenticity in supportive communication.

\subsection{Conclusion}
Online health communities (OHCs) like CaringBridge are predicated upon a specialized set of user needs, supporting millions of people facing challenging health conditions. Like many mainstream social media platforms, CaringBridge offers mechanisms for visitors to respond to posts including commenting and the paralinguistic affordance of single-click ``Liking.'' In 2016, Facebook's reaction launch introduced reacting as an \textit{emotion}-based PDA. By evaluating the similar launch of CaringBridge reactions in 2021, in which~\raisebox{-.5\mydepth}{\includegraphics[height=\myheight]{reactions/amp.jpg}} was extended into a set of symbols including Prayer~\raisebox{-.5\mydepth}{\includegraphics[height=\myheight]{reactions/pray.jpg}}, Happy~\raisebox{-.5\mydepth}{\includegraphics[height=\myheight]{reactions/happy.jpg}}, and Sad~\raisebox{-.5\mydepth}{\includegraphics[height=\myheight]{reactions/sad.jpg}}, our work finds that Facebook's reaction-style may not be best for CaringBridge in two major ways: (1) First, we provide evidence of specialized meaningfulness associated with CaringBridge's \textit{single-option, single-click} ~\raisebox{-.5\mydepth}{\includegraphics[height=\myheight]{reactions/amp.jpg}} reaction affordance: it can be universally applied to any Journal update as an expression of acknowledgement and support. The simplicity and intuitiveness of the feature may make it conceptually preferable to the \textit{limited-bar} reaction style, which introduces ambiguity and cognitive complexity. (2) Second, we also find that over 90\% of users are accepting of the new set of reactions, or want even more reactions--even if conceptually, there were benefits to the single-click. Rather than returning to single-click, CaringBridge could refine its set of options. In particular, an \textit{emotion}-based reaction style appears to introduce discomforts and confusions in this health-critical context. We suggest that a health-promoting \textit{gesture}-based reaction set may better address the CMC needs of authors and visitors, such as: Heart~\raisebox{-.5\mydepth}{\includegraphics[height=\myheight]{reactions/amp.jpg}}, and Prayer~\raisebox{-.5\mydepth}{\includegraphics[height=\myheight]{reactions/pray.jpg}}, alongside custom-branded graphics similar to 
Hug~\raisebox{-.5\mydepth}{\includegraphics[height=\myheight]{emoji/hug.png}}~, Flowers~\raisebox{-.5\mydepth}{\includegraphics[height=\myheight]{emoji/bouquet_1f490.png}}~, Candle~\raisebox{-.5\mydepth}{\includegraphics[height=\myheight]{emoji/candle.png}}, and Clap~\raisebox{-.5\mydepth}{\includegraphics[height=\myheight]{emoji/clapping-hands_1f44f.png}}---ideally with non-yellow skin tones \textit{and} the ability to opt-out of any of these.

Our work contributes to the growing body of knowledge on emoji, reactions, and the many ways in which we can design health-support technologies to become better at aligning the needs of patients and caregivers with the capacities of their support networks to meet them. What works on mainstream platforms does not always work for niche platforms, but it can nonetheless inspire meaningful technological innovation. Future designers can use the insights and design strategies suggested by this work to improve health platforms, and future researchers can benefit from the research directions and questions we have provided for future work.

\begin{acks}
We want to thank our collaborators at CaringBridge who helped us to run our survey early in 2021, as well as all of the CaringBridge users who volunteered their time to participate in our survey and interviews. We also want to thank Hai Nguyen, a software engineer at CaringBridge, who attended the first author's PhD dissertation defense and asked a question about emoji reactions in relation to spiritual support (since CaringBridge was planning to launch new reactions soon)---that question initiated this study. \raisebox{-.5\mydepth}{\includegraphics[height=\myheight]{reactions/happy.jpg}} Finally, the first author thanks Brian Keegan and Chenhao Tan for enabling her to pursue this research during her postdoctoral appointment supported by NSF Award \#1910225.
\end{acks}

\bibliographystyle{ACM-Reference-Format}
\bibliography{0_main}

\appendix

\section{Data Examples}\label{sec:dataexamples}

\begin{table}[H]
\begin{tabular}{ p{7cm} | p{7cm} }
\toprule
AGAINST &
  FOR \\ \midrule \midrule
We chose CaringBridge because it's not social media. I think a lot gets lost in reactions or different emojis. It makes it not as meaningful. Having additional emojis and making CaringBridge more like Facebook would seriously make me consider not using it. (S563) &
  I enjoy good imogees because I'm more of a visual person than a verbal person and expressions convey quite a bit of meaning with very little space/time required.  I think they are a great idea and allow a form of short hand and/or a re-enforcement of an expressed thought. (S288) \\ \midrule
No more of these please. We are becoming an increasingly illiterate society. Surely people can write “I am so sorry for your troubles” instead of poking a cartoon face. (S621) &
  I feel that emojis allow people a way to express themselves that may not be correctly seen with words. For instance, I am a rather sarcastic person. Unless you know me, you likely won't feel that in my writing. (S321) \\ \midrule
I would discourage too many reactions being introduced. Part of why caring bridge is nice is because it is a platform for communication that isn't obstructed by all the noise other forms of social media platforms have. It is straight forward: information, and simple messages/reactions. (S97) &
  I think it would be beneficial to have the full array of emojis available to the user so that they can express themselves through emojis when words are challenging. I tend to use \raisebox{-\mydepth}{\includegraphics[height=\myheight]{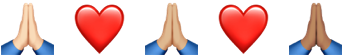}} when I don’t have the words for a response to people’s incredible caring. (S406) \\
 \bottomrule
\end{tabular}
\caption{Examples of Arguments Against and For Adding Additional Reactions to CaringBridge}
\label{tab:debate}
\end{table}

\section{Subgroup Analysis Comparing Users who Had \& Had Not Used New Reactions}\label{sec:subgroups}

404 survey respondents indicated they had seen or used the new reactions, while the remaining half indicated they had not. We use the term \textit{``exposed''} to refer to participants who had already seen the new reactions, and \textit{``non-exposed''} to those who had not. 

Exposed respondents were more likely to be visitors than authors---compared to non-exposed respondents (47.3\% vs 60.9\% authors, $t=-3.91$, $p<0.001$). Demographics were otherwise similar between the two groups.
To assess whether previous exposure to the reactions affected our results, we conducted uncorrected t-tests comparing exposed and non-exposed respondents for each of the quantitative results reported in the paper. Only one of these tests had a notably small p-value (bolded below). Result comparison:

\begin{itemize}
\item Interview opt-in: A similar proportion of exposed vs. non-exposed respondents volunteered for interviews (41.6\% vs 47.8\% volunteered, $t=-1.77$, $p=0.077$).
\item Visitor reaction rationale (sec.~\ref{sec:rationale}): A similar proportion of exposed vs non-exposed visitors often or always ``don't know what to write in a comment'' (50.2\% vs 51.3\%, $t=-0.20$, $p=0.845$) and use a reaction when they don't know what to write (53.5\% vs 55.7\%, $t=-0.41$, $p=0.678$). 
\item Heart perceptions (sec.~\ref{sec:perceptions_of_heart}): A similar proportion of exposed vs non-exposed respondents agree that ``Heart is unique to CaringBridge'' (78.7\% vs 80.0\%, $t=-0.43$, $p=0.664$).
\item Heart perceptions, visitor (sec.~\ref{sec:perceptions_of_heart}): Exposed vs non-exposed visitors agree the Heart is a way of saying ``I support you'' (76.5\% vs 82.3\%, $t=-1.34$, $p=0.180$).
\item Heart perceptions, author (sec.~\ref{sec:perceptions_of_heart}): Exposed vs non-exposed authors both feel personally supported receiving a Heart (85.9\% vs 85.4\%, $t=0.15$, $p=0.883$), but non-exposed authors were more likely to indicate that they would rather receive a Heart than nothing (73.3\% vs 81.7\%, $t=-2.11$, $\mathbf{p=0.035}$).
\item Reactions expansion (sec.~\ref{sec:rq2_results}): A similar proportion of respondents indicated a preference for more reactions on CaringBridge (49.0\% vs 44.8\%, $t=1.20$, $p=0.231$) and that the new set of four is sufficient (44.6\% vs 45.8\%, $t=-0.35$, $p=0.724$).
\item Reactions versus comments (sec.~\ref{sec:reactionsvscomments}): A similar proportion of authors responded that receiving comments is more important to them than receiving reactions (48.2\% vs 47.2\%, $t=0.21$, $p=0.834$).

\end{itemize}

We find small overall differences between exposed and non-exposed survey respondents.
Therefore, throughout the paper, we report quantitative results including all 808 participants.

\section{Future research questions}\label{ref:futureRQs}
The qualitative results presented in this paper suggest the following research questions, which can be addressed quantitatively in future analyses of behavioral data on CaringBride or other social media platforms:

\begin{enumerate}
    \item Does adding new reactions increase the \textit{overall} number of interactions on the platform? (Sec.~\ref{sec:rationale} and~\ref{sec:expectingresponse}) %
    \item How does the introduction of new reactions impact commenting behaviors? If new reactions \textit{decrease} commenting, are \raisebox{-\mydepth}{\includegraphics[height=\myheight]{reactions/pray.jpg}} reactions replacing comments containing exclusively series of emoji or simple expressions like ``praying for you'', or are they impacting other types of comments, as well? (Sec.~\ref{sec:rationale} and ~\ref{sec:reactionsvscomments})
    \item How do textual features of journals relate to reactions (or combinations of reactions) received? (For example, do people empathize by using sad with sad updates, happy with happy updates?) Can we predict what reactions an update will receive based on the text? (Sec.~\ref{sec:functionalambiguity})
    \item Does the intensity of language used, or higher inclusion of ``medical-related'' words, correlate with higher volume of reactions and comments? Do these considerations impact the proportion of reactions to comments? (Sec.~\ref{sec:reactionsvscomments})
    \item Is there an ``optimal ratio'' of reactions to comments that contributes to better community survival time? (Sec.~\ref{sec:reactionsvscomments})
    \item Are people who are relationally closer (Sec.~\ref{sec:socialcloseness}) or older (Sec.~\ref{sec:personalcharacteristics}) more likely to leave comments? Are self-identified Christians more likely to use Prayer (Sec.~\ref{sec:personalcharacteristics})?
    \item Are there identifiable patterns in reacting \textit{v.s.} commenting behaviors? Are there specific ``reactor'' or ``commenter'' user types, or are there population-level patterns? (Sec.~\ref{sec:reactionsvscomments} and~\ref{sec:personalcharacteristics}) %
    \item Are latecomers more likely to react than to comment, because they may not have anything new to add that hasn't been written already? (Sec.~\ref{sec:socialdynamics})
    \item Are later reactors more likely to leave the same reactions that have already been left? (Sec.~\ref{sec:socialdynamics}) %
\end{enumerate}

\received{July 2022}
\received[revised]{January 2023}
\received[accepted]{March 2023}

\end{document}